\documentclass[10pt]{article}

\usepackage[utf8]{inputenc}
\usepackage[english]{babel}
\usepackage[toc,page]{appendix}

\usepackage{amsmath,amsfonts,amssymb,amsthm}
\usepackage{graphicx}
\usepackage{float}
\usepackage{hyperref}
\usepackage{dsfont}
\usepackage{subcaption}
\usepackage[format=plain,font=small,textfont=it]{caption}
\usepackage[dvipsnames]{xcolor}
\usepackage{multirow} 
\usepackage{bookmark}
\usepackage{mathrsfs}
\usepackage{algorithm}
\usepackage[noend]{algpseudocode}

\graphicspath{{img/}}

\newcommand{\C}{\mathbb{C}}

\newcommand{\dd}{\mathrm{d}}
\newcommand{\Wthree}[6]{\left(\begin{array}{ccc} #1 & #2 & #3 \\ #4 & #5 & #6 \end{array}\right)}
\newcommand{\Wfour}[9]{\left(\begin{array}{cccc} #1 & #2 & #3 & #4 \\ #5 & #6 & #7 & #8 \end{array}\right)^{(#9)}}
\newcommand{\Wsix}[6]{\left \{ \begin{array}{ccc} #1 & #2 & #3 \\ #4 & #5 & #6 \end{array}\right \} }

\begin{document}

\title{Summing bulk quantum numbers with Monte Carlo\\ in spin foam theories}
% Keywords: Monte Carlo, EPRL Spin foam theory, bulk degrees of freedom, many vertices
%The title has to be as short as possible

\author{\Large{Pietro Don\`a${}^{a,b}$\footnote{pdona@uwo.ca}, \ } \Large{Pietropaolo Frisoni${}^{b}$\footnote{pfrisoni@uwo.ca} \ } 
\smallskip \\ 
\small{\textit{${}^a$ Center for Space, Time and the Quantum, 13288 Marseille, France}}\\
\small{\textit{${}^b$ Department of Physics and Astronomy, University of Western Ontario, London, ON N6A 5B7, Canada}}}

\date{}

\maketitle

\begin{abstract}
We introduce a strategy to compute EPRL spin foam amplitudes with many internal faces numerically. We work with \texttt{sl2cfoam-next}, the state-of-the-art framework to numerically evaluate spin foam transition amplitudes. We find that uniform sampling Monte Carlo is exceptionally effective in approximating the sum over internal quantum numbers of a spin foam amplitude, considerably reducing the computational resources necessary. We apply it to compute large volume divergences of the theory and find surprising numerical evidence that the EPRL vertex renormalization amplitude is instead finite. 
\end{abstract}

%%%%%%%%%%%%%%%%%%%%%%%%%%%%%%%%%%%%%%%%%%%%%%%%%%%%%%%%%%%%%%%%%%%%%%%%%%%%%%%%%%%%%%%%%%

\section{Introduction}
\label{sec:intro}
Spin foam theory is the Lorentz covariant version of loop quantum gravity (LQG) and provides a tentative background independent path integral for gravity. It gives dynamics to LQG kinematical states defining transition amplitudes between spin network states \cite{Rovelli:2014ssa, Perez:2012wv}. 
The most promising spin foam theory is the EPRL-FK model \cite{Engle:2007wy, Freidel:2007py}. Various generalizations include the extension to general triangulations \cite{Kaminski:2009fm}, the inclusion of a cosmological constant \cite{Han:2011aa,Han:2021tzw}, and boundary with different signature \cite{Conrady:2010kc}.
These theories promisingly connect with discrete general relativity in the double limit of finer discretization and large areas \cite{Barrett:2009mw, Dona:2020yao, Dona:2020tvv, Han:2021kll, Engle:2021xfs,Dona:2022hgr}. 

\medskip 

The field has recently undergone an explosion of numerical methods, providing new tools to address many open questions of the theory. We can compute expectation values and fluctuations of operators in the large spins regime using the complex saddle point analysis and the integration on a Lefschetz thimble using Markov chain Monte Carlo \cite{Han:2021kll, Han:2020npv,Han:2023cen}. It is possible to verify that Regge geometries emerge in the large-scale and small Immirzi parameter regime. A similar result can also be obtained from effective spin foam models \cite{Asante:2020iwm,Asante:2021zzh}. 
Finally, \texttt{sl2cfoam} (and its latest iteration \texttt{sl2cfoam-next}) is an open source framework to compute EPRL spin foam amplitudes \cite{Dona:2018nev,Gozzini:2021kbt}. It is based on a divide-and-conquer strategy, and a booster decomposition of the vertex amplitude \cite{Speziale:2016axj}. The library was already employed to explore the large quantum numbers regime \cite{Gozzini:2021kbt, Dona:2019dkf, Dona:2022dxs}, the infrared divergences of the theory \cite{Frisoni:2021uwx,Dona:2022vyh}, the black-to-white hole transition \cite{Frisoni:2023agk}, and correlations in the early universe \cite{Frisoni:2022urv}. Very recently, a hybrid approach taking advantage of all the available techniques was also proposed \cite{Asante:2022lnp}.

\medskip 

This work overcomes one of the principal limitations of \texttt{sl2cfoam-next}. The library provides an optimized and efficient framework to compute all the constituents of a spin foam transition amplitude. Nevertheless, numerically computing spin foam amplitudes with many internal faces is prohibitively taxing. 
There are way too many objects to compute as their number scales exponentially with the number of internal faces. We overcome this problem by evaluating the sums over the internal quantum numbers using statistical frameworks. We explore the possibility of using uniform sampling Monte Carlo and find it surprisingly effective. We can compute amplitudes with slightly better than $1\%$ precision by considering a sampling five orders of magnitude smaller than the total amount of terms of the sum. 

\medskip 

We apply this novel technique to compute the melonic self-energy and the vertex renormalization amplitude in the $SU(2)$ BF and EPRL theories. These two amplitudes are the perfect laboratory to test the effectiveness of Monte Carlo as they possess many internal faces. These spin foam amplitudes are believed to be divergent, and their renormalization is crucial to define the continuum limit of the EPRL theory. In the case of the melonic self-energy we find excellent agreement with the numerical results in the literature \cite{Frisoni:2021dlk, Frisoni:2021uwx}, which do not use stochastic methods. This work contains the first computation of the Lorentzian EPRL vertex renormalization amplitude. We surprisingly find numerical evidence for its convergence. 

\medskip

The scripts we use to compute the amplitudes and the notebooks to analyze the data are publicly available at the repository \cite{PaperRepo}. We perform most of our calculations on the \textit{Narval} cluster of the Digital Research Alliance of Canada. 

%%%%%%%%%%%%%%%%%%%%%%%%%%%%%%%%%%%%%%%%%%%%%%%%%%%%%%%%%%%%%%%%%%%%%%%%%%%%%%%%%%%%%%%%%%
\section{Spin foam transition amplitudes}
\label{sec:spinfoam}
We write a spin foam transition amplitude starting from a 2-complex $\Delta$ of a simplicial triangulation of the space-time manifold decorated with LQG quantum numbers. Each face is colored with a spin $j_f$ and each edge with an intertwiner $i_e$.

The spin foam transition amplitude $A_\Delta$ is the product of local fundamental amplitudes: a face amplitude $A_f(j_f)$, an edge amplitude $A_e(i_e)$, and a vertex amplitude $A_v \left(j_f, \  i_e\right)$. Finally, we sum over all the possible quantum numbers associated with the bulk of the 2-complex
\begin{equation}
\label{eq:transition_amplitude}
A_{\Delta} = \sum_{j_f=0}^{\infty} \sum_{i_e}  \prod_f A_f(j_f) \prod_e A_e(i_e) \prod_v A_v \left(j_f, \  i_e\right) \ .
\end{equation}

This work focuses on two spin foam theories: the topological BF $SU(2)$ model and the Lorentzian EPRL model. We introduce them here schematically and report their detailed definition in Appendix~\ref{app:vertexdetails}. We refer to reviews \cite{Perez:2012wv} or books \cite{Rovelli:2014ssa} for a more complete and pedagogical introduction. We use the same notation for the vertex amplitudes in the two models. It is convenient to avoid overburdening the notation and not repeat the same equations twice. We will stress the difference between the two models if necessary. The vertex amplitude for the topological model is defined as
\begin{equation}
\label{eq:vertexamplitudeBF}
    A_v \left(j_f, \  i_e\right) =   \{ 15 j\} (j_f; i_e)  \ ,
\end{equation}
where the $\{ 15 j\} (j_f; i_e)$ is a $SU(2)$ invariant depending on the ten spins and five intertwiners coloring the spin foam vertex. 
We work with the booster functions decomposition of the Lorentzian EPRL spin foam model introduced in  \cite{Speziale:2016axj}. In this form, the EPRL vertex amplitude is a superposition of $\{15j\}$ symbols weighted by booster functions $B_4^\gamma$.
\begin{equation}
\label{eq:vertexamplitude}
    A_v \left(j_f, \  i_e\right) =   \sum_{l_f=j_f}^\infty \sum_{k_e} \{ 15 j\} (j_f,l_f)  \prod_{e = 2}^{5}  (2k_{e}+1) B_4^\gamma \left( l_f, j_f ; i_{e} ,k_{e} \right)   \ .
\end{equation}
The presence of the booster functions is the striking difference between the amplitudes \eqref{eq:vertexamplitudeBF} and \eqref{eq:vertexamplitude}. They encode the imposition of the simplicity constraints and the explicit dependence of the theory from the Immirzi parameter $\gamma$. They possess a compelling geometrical interpretation of boosted tetrahedra \cite{Dona:2020xzv}.
The edge and face amplitudes are fixed, requiring the correct composition of spin foam amplitudes \cite{Bianchi:2010fj}.
\begin{equation}
    \label{eq:edgefaceamplitude}
    A_e \left(i_e\right) = 2 i_e +1\ , \qquad \text{and} \qquad
    A_f \left(j_f\right) = 2 j_f +1 \ .
\end{equation}
Depending on the details of the 2-complex could be necessary to also multiply by some extra phase and edge-related $SU(2)$ invariants depending on the spin and intertwiners quantum numbers. They result from our decision to work with a specific recoupling scheme in the vertex amplitudes. We refer to the review \cite{Dona:2022dxs} for a step-by-step guide on computing them. 

We perform all the numerical calculations using \texttt{sl2cfoam-next}, the state-of-the-art code, to compute spin foam amplitudes with a computer. The library is open source and written in \texttt{C}. It is based on the booster decomposition of the EPRL vertex amplitude, optimizing the available computational resources. We refer to the original paper \cite{Gozzini:2021kbt}, the review \cite{Dona:2022dxs} or the book chapter \cite{Dona:2023hpc} for a detailed description. 

One of the main ingredients we mention here is the introduction of a homogeneous truncation parameter $\Delta l$ to approximate the unbounded convergent sums over the virtual spins $l_f$ in \eqref{eq:vertexamplitude}. 
\begin{equation}
     \sum_{l_f=j_f}^\infty \qquad \longrightarrow \qquad  \sum_{l_f=j_f}^{j_f + \Delta l} \ .
\end{equation}
Despite the notation, we emphasize that the truncation parameter $\Delta l$ is independent of the 2-complex $\Delta$.

%%%%%%%%%%%%%%%%%%%%%%%%%%%%%%%%%%%%%%%%%%%%%%%%%%%%%%%%%%%%%%%%%%%%%%%%%%%%%%%%%%%%%%%%%%

\section{Summing bulk degrees of freedom with Monte Carlo}
\label{sec:sumMC}
The library \texttt{sl2cfoam-next} \cite{Gozzini:2021kbt} computes EPRL vertex amplitudes \eqref{eq:vertexamplitude} very fast and efficiently\footnote{It can evaluate the topological BF $SU(2)$ vertex amplitude too.}. Unfortunately, it is not enough to compute a general spin foam amplitude with many vertices and internal faces. The number of vertex amplitudes we have to calculate, assemble, and sum grows exponentially with the number of bulk faces. We can convince ourselves this is a severe problem with a back-of-the-envelope calculation. Imagine you want to compute an amplitude with $F$ internal faces, and all the spins associated with the inner faces have some characteristic value $J$. To calculate the amplitude, we must loop through all $(2J+1)^F$ possible values that the internal spins can assume and compute all the vertex amplitudes. Let's assume, optimistically, that we need just $1\mu s$ of CPU time to obtain them (the actual time is orders of magnitude larger). Suppose we want to calculate an amplitude with $10$ internal faces and average spins of order $10$. To perform this calculation, we need approximately $21^{10} \mu s \approx 6$ months of CPU time, which is a lot of time considering our modest requirements and optimistic hypothesis. 

\medskip

We overcome this problem using Monte Carlo to perform the sum over the bulk spins. We rewrite the spin foam amplitude \eqref{eq:transition_amplitude} in terms of partial amplitudes
\begin{equation}
\label{eq:restricted_amplitude}
A_{\Delta} = \sum_{j_f=0}^{\infty} a (j_f) \qquad \text{with} \qquad  a(j_f) =  \sum_{i_e}  \prod_f A_f(j_f) \prod_e A_e(i_e) \prod_v A_v \left(j_f, \  i_e\right) \ .
\end{equation}
We include the sum over the bulk intertwiners in the partial amplitudes $a(j_f)$. Those sums are always finite for fixed bulk spins, and we perform them leveraging the tensorial structure of \texttt{sl2cfoam-next} without resorting to Monte Carlo methods. We are omitting the evident dependence from the 2-complex $\Delta$ of the partial amplitude. 

The partial amplitude vanishes if some spins $j_f$ do not satisfy triangular inequalities\footnote{
For example, if $j_1$, $j_2$, $j_3$, and $j_4$ are the four spins associated with the faces contained in an edge of $\Delta$ and
\begin{equation*}
     \mathrm{Max}(|j_1-j_2|,|j_3-j_4|)> \mathrm{Min}(j_1+j_2,j_3+j_4) \Rightarrow a_\Delta (j_f) =0 
\end{equation*}
then the set of intertwiners associated with that edge is empty, and the partial amplitude trivially vanishes.
}.
We restrict the sum over the bulk spins in \eqref{eq:restricted_amplitude} to the set of spins satisfying triangular inequalities that we indicate as $\mathcal{I}_\Delta$. In this way, we eliminate the majority of trivially vanishing partial amplitudes. This step is convenient to obtain an efficient Monte Carlo amplitude estimate. 

Generally, $\mathcal{I}_\Delta$ is unbounded, therefore is not possible to directly apply Monte Carlo to estimate the amplitude. We circumvent this limitation subdividing $\mathcal{I}_\Delta$ into layers $\mathcal{J}_{k}$. 
\begin{equation}
\mathcal{J}_{k} = \left\lbrace j_f \in \mathcal{I}_{\Delta} \vert \max j_f = k \right\rbrace \ .
\end{equation}
By definition, each layer is finite. Different layers do not overlap $\mathcal{J}_{k} \cap \mathcal{J}_{k'} = \emptyset$, and the union of all of them forms the original set $\mathcal{I}_{\Delta} = \bigcup_{k=0}^{\infty} \mathcal{J}_{k}$. We reorganize the spin foam amplitude as a sum over layers of layer amplitudes
\begin{equation}
\label{eq:layered_amplitude}
A_{\Delta} = \sum_{k=0}^{\infty} S_k \qquad \text{with} \qquad S_k = \sum_{j_f \in \mathcal{J}_{k} }a(j_f) \ .
\end{equation}
Each layer amplitude $S_k$ is defined as a sum with a finite number of terms (that, however, grows rapidly with $k$). We can approximate them with Monte Carlo $S_k^{mc}$ using the procedure described in Appendix~\ref{app:MCSums} with a fixed amount of samples $N_{mc}$. In general, the number of layers is infinite\footnote{In some exceptional cases, the layers are all empty from a particular value of $k$ forward because of the triangular inequalities involving both bulk and boundary spins. It is the case of the $\Delta_3$, and the $\Delta_4$ triangulations studied numerically in \cite{Dona:2020tvv,Dona:2022dxs}}. We cutoff the sum over the layer to a maximum layer $K$. This prescription is equivalent to introducing a homogeneous cutoff $K$ to all the bulk spins. The Monte Carlo approximation of a spin foam amplitude with a cutoff $K$ is given by 
\begin{equation}
\label{eq:monte_carlo_amplitude}
A_{\Delta}^{mc} (K) = \sum_{k=0}^{K} S_k^{mc} \ .
\end{equation}

The calculation of $S_k^{mc}$ requires a discrete random uniform probability distribution over the layer $\mathcal{J}_k$. In principle, we could map the layer in an interval of integers, define a uniform distribution there and map it back to $\mathcal{J}_k$. This prescription is very unpractical. We prefer to define the uniform distribution in an alternative way. 

We extract one real number from a continuous uniform distribution in $[0,k+0.5]$ for each bulk face. We floor them to half-integers and we check if they belong to the layer $\mathcal{J}_{k}$. If they do, we accept them as a random sample of the layer. If they do not, we discard them and repeat the procedure. We summarize this procedure in the flowchart~\ref{algo:numericalsampling}. 
\begin{algorithm}[H]
\caption{Random sampling of spins in $\mathcal{J}_k$}\label{algo:numericalsampling}
\begin{algorithmic}[1]
\Procedure{RandomSample}{$k$,$\mathcal{J}_k$}  
\While{\textbf{true}} \Comment{repeat until we find a good candidate}
\State $j_{f}\gets$ extract a real number from a uniform distribution in $[0,k+0.5]$ for each face
\State $j_{f}\gets$ floor them to half-integers and interpret them as spins
\If{all spins $j_{f}$ are smaller than $k$} 
\State \textbf{continue}
\EndIf 
\If{any spin $j_{f}$ do not satisfy triangular inequalities of $\mathcal{J}_k$}
\State \textbf{continue}
\EndIf 
\State \textbf{return} $j_f$ \Comment{the spins belong to the layer $\mathcal{J}_k$}
\EndWhile
\EndProcedure
\end{algorithmic}
\end{algorithm}
In the application we present in Section~\ref{sec:MelonBF}, we perform an explicit test to show that the samples extracted with this algorithm are uniformly distributed in $\mathcal{J}_{k}$.

We acknowledge that Algorithm~\ref{algo:numericalsampling} is not optimal. To scale it up to more complicated amplitudes, we must improve it considerably. We leave this task to future work. Since they share the edge structure, the sampling Algorithm \ref{algo:numericalsampling} is the same for both spin foam models we study.

\medskip

We conclude this section by showing a pseudocode representation (Algorithm~\ref{algo:numerical_MC_BF}) of the \texttt{Julia} scripts that implement the Monte Carlo estimate of the spin foam amplitude \eqref{eq:monte_carlo_amplitude}. The full Julia scripts are available in the repository \cite{PaperRepo}.

\begin{algorithm}[H]
\caption{Scheme of Monte Carlo estimate of the spin foam amplitude}\label{algo:numerical_MC_BF}
\begin{algorithmic}[1]
\For{$k = 0.5 \ , 1 \dots K$} \Comment{for each layer}
\State extract $N_{mc}$ samples of the layer using \Call{RandomSample}{$k$,$\mathcal{J}_k$}
\State store them in memory
\EndFor
\State
\For{$k = 0.5 \ , 1 \dots K$} \Comment{for each layer}
\State load the bulk spins samples from memory 
\For{$n = 1 \dots N_{mc}$} \Comment{for each sample}
\State compute the partial amplitude $a (j_{f})$
\EndFor
\State sum the partial amplitudes and save the layer amplitude $S_k$
\State compute the amplitude $A_{\Delta}^{mc} (k)$
 summing the layer $k$ to the previous ones 
 \State store the amplitude
\EndFor
\end{algorithmic}
\end{algorithm}

In calculating an EPRL spin foam amplitude, the partial amplitude also depends on the truncation parameter $\Delta l$. We fix the truncation parameter once and for all and store the amplitude for every $\Delta l$.

%%%%%%%%%%%%%%%%%%%%%%%%%%%%%%%%%%%%%%%%%%%%%%%%%%%%%%%%%%%%%%%%%%%%%%%%%%%%%%%%%%%%%%%%%%
\section{Applications to the melonic self-energy and vertex renormalization diagrams}
\label{sec:applications}
We test the effectiveness of the Monte Carlo framework described in Section~\ref{sec:sumMC} by computing four spin foam amplitudes.
We focus on the melonic self-energy amplitude and the vertex renormalization (or $5-1$ Pachner move) amplitude with the topological BF $SU(2)$ and the EPRL model. These diagrams are essential for studying the infrared divergences of spin foam theories and their continuum limit. 

\begin{figure}[H]
    \centering
    \raisebox{-0.5\height}{\includegraphics[scale=0.5]{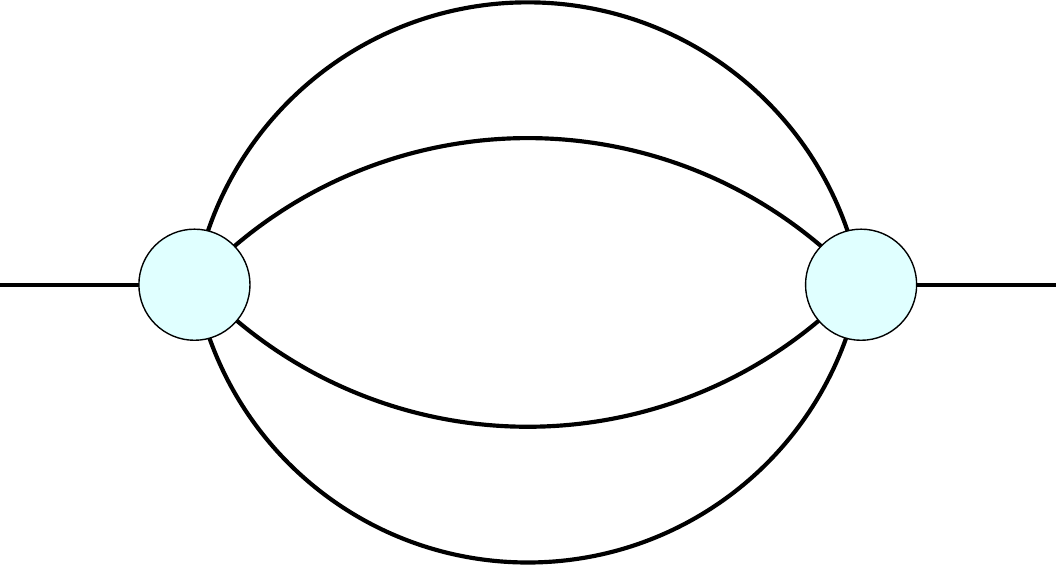}}
    \hspace{2cm}
    \raisebox{-0.5\height}{\includegraphics[scale=0.5]{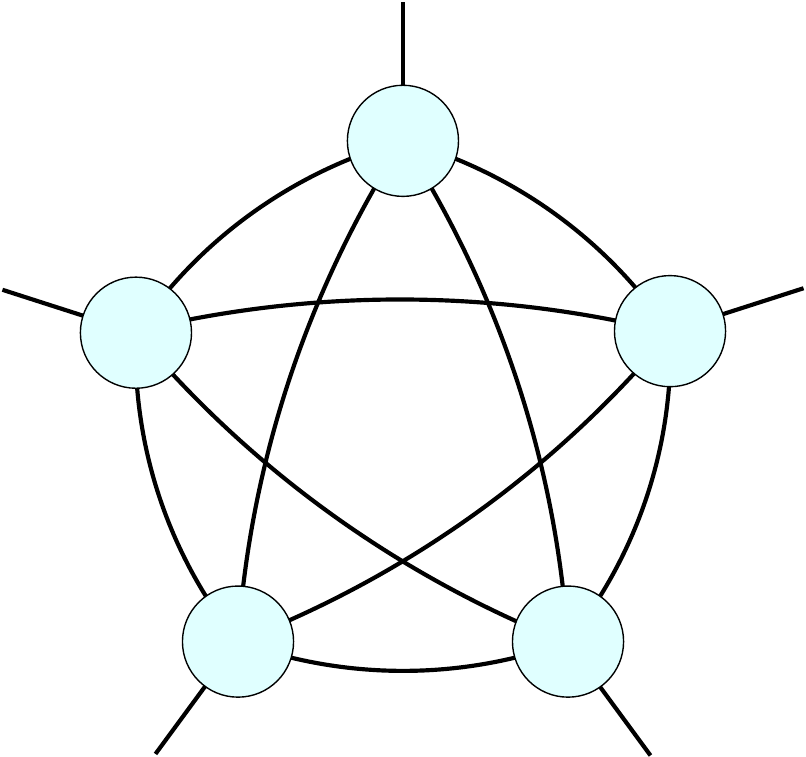}}
    \caption{Schematic representation of the 2-complexes of the self-energy spin foam diagram (left) and the vertex renormalization spin foam diagram (right).}
    \label{fig:2-complexes}
\end{figure}

The melonic self-energy diagram comprises two vertices, two boundary edges (one for each vertex), and four bulk ones connecting the two vertices, four boundary faces, and six bulk faces. We report in Figure~\ref{fig:2-complexes} a schematic representation of the 2-complex, and we refer to Appendix~\ref{app:details_diagrams} for the detailed routing diagram. 

To simplify the numerical calculation, we consider very symmetric boundary data where the boundary spins $j_b$ are all the same and all the boundary intertwiners $i_b$ are also all the same. The spin foam transition amplitude with a homogeneous cutoff on all the spins associated with bulk faces is 
\begin{equation}
    \label{eq:melon}
    \begin{split}
     A_{se} \left(j_b, \  i_b; \, K\right) = \sum_{j_f = 0}^{K} \left(\prod_{f=1}^{6} (2 j_f +1)\right) \sum_{i_e} \left(\prod_{e=1}^{4} (2 i_e +1)\right)& A_v \left(  j_b, j_b, j_b, j_b, j_{1}, j_{2}, j_{3}, j_{4}, j_{5}, j_{6}; \ i_b , i_1, i_2, i_3, i_4\right) \\[-1.2em]
     &  A_v \left(  j_b, j_b, j_b, j_b, j_{1}, j_{2}, j_{3}, j_{4}, j_{5}, j_{6}; \ i_b , i_4, i_3, i_2, i_1\right) \ .
    \end{split}
\end{equation}
The amplitude \eqref{eq:melon} is the same for both spin foam models that differ by the vertex amplitudes $A_v$: \eqref{eq:vertexamplitudeBF} for BF SU(2) and \eqref{eq:vertexamplitude} for EPRL. The melonic self-energy diagram with the BF $SU(2)$ topological model can be evaluated analytically and numerically \cite{Frisoni:2021uwx, Dona:2018pxq, Frisoni:2021dlk}. We know it is divergent. The divergence is due to redundant $SU(2)$ delta functions that indicate some residual gauge freedom in the path integral \cite{Freidel:2002dw} and can be dealt with by gauge fixing appropriately. The same amplitude with the EPRL model has been studied analytically \cite{Riello:2013bzw}, with a hybrid calculation \cite{Dona:2018pxq}, and, recently, numerically \cite{Frisoni:2021uwx, Frisoni:2021dlk}. The amplitude is divergent, and there are strong indications that it diverges linearly in the cutoff. We are revisiting this amplitude as a control for the Monte Carlo technique we introduce. In fact, the self-energy has a relatively small number of internal faces, so the computation is still possible even without using Monte Carlo. Reproducing known results allows us to evaluate the choices of the framework. 

\medskip

The vertex renormalization diagram contains five vertices, five boundary edges (one for each vertex), and ten bulk ones connecting all couples of vertices, ten boundary faces, and ten bulk faces. We report in Figure~\ref{fig:2-complexes} a schematic representation of the 2-complex, and we refer to Appendix~\ref{app:details_diagrams} for the detailed routing diagram. 

Also in this diagram, we simplify the numerical calculation by taking symmetric boundary data with all equal boundary spins $j_b$ and boundary intertwiners $i_b$. If we put a homogeneous cutoff $K$ on the sums over the bulk spins, the amplitude reads 
\begin{equation}
    \label{eq:vertex}
    \begin{split}
     A_{vr} \left(j_b, \  i_b; \, K\right) = \sum_{j_f = 0}^{K} \left(\prod_{f=1}^{10} (2 j_f +1) \right) \sum_{i_e} \left(\prod_{e=1}^{15} (2 i_e +1)\right)&A_v(j_b, j_b, j_b, j_b, j_1, j_2, j_3, j_4, j_5, j_6 ; \ i_b, i_4 ,i_{11} ,i_{12} ,i_{2} ) \\[-1.2em]
&A_v(j_b, j_b, j_b, j_b, j_7, j_8, j_1, j_9, j_2, j_3 ; \ i_b, i_6 ,i_{13} ,i_{14} ,i_4 ) \\
&A_v(j_b, j_b, j_b, j_b, j_{10}, j_4, j_7, j_5, j_8, j_1 ; \ i_b, i_8 ,i_{15} ,i_{11} ,i_6 ) \\
&A_v(j_b, j_b, j_b, j_b, j_6, j_9, j_{10}, j_2, j_4, j_7 ; \ i_b, i_{10} ,i_{12} ,i_{14} ,i_{15} ) \\
&A_v(j_b, j_b, j_b, j_b, j_3, j_5, j_6, j_8, j_9, j_{10} ; \ i_b, i_2 ,i_{13} ,i_{15} ,i_{10}) \\
&\{6j\}(j_b,j_3,i_1,j_5, j_6, i_2)\{6j\}(j_b,j_1,i_3,j_2, j_3, i_4)  \\ 
&\{6j\}(j_b,j_7,i_5,j_8, j_1, i_6)\{6j\}(j_b,j_{10},i_7,j_4, j_7, i_{8})  \\
&\{6j\}(j_b,j_6,i_8,j_9, j_{10}, i_{10})  (-1)^{\chi}  \ ,
    \end{split}
\end{equation}
% This is the correspondence with spins and intertwiners used in the code:
% SPINS
% jpink=j1   jblue=j2   jbrightgreen=j3   jpurple=j4   jgrassgreen=j5   jred=j6   jbrown=j7   jdarkgreen=j8   jorange=j9   jviolet=j10
% INTERTWINERS
% rBCl=i1   rBCr=i2
% rABl=i3   rABr=i4
% rAEl=i5   rAEr=i6
% rbl=i7   rbr=i8
% rCDl=i9   rCDr=i10
% rIul=i11   rIur=i12
% rIu =i13  
% rIbl=i14   rIbr=i15
% let's keep this comment
where $\{6j\}$ are $SU(2)$ invariants that we define in Appendix~\ref{app:vertexdetails}. The phase in \eqref{eq:vertex} reduces to:
\begin{equation}
    \chi = \sum_{k = 1}^{10} j_k + \sum_{k = 11}^{15} i_k \ .
\end{equation}
The form of the amplitude is convoluted because we want to use the same intertwiner recoupling scheme in all the vertices. This is necessary to perform the numerical calculations efficiently, as \texttt{sl2cfoam-next} implements only a specific vertex amplitude \eqref{eq:EPRL_vertex_amplitude}. Again, the amplitude \eqref{eq:vertex} is the same for both spin foam models that differ by the vertex amplitudes. 
The vertex renormalization diagram can be evaluated analytically with the BF $SU(2)$ topological model integrating explicitly the group functions in the holonomy representation of the amplitude \cite{Dona:2018pxq}. This amplitude was already studied with the Euclidean EPRL model in \cite{Perini:2008pd}, finding a logarithmic divergence. A numerical calculation of the amplitude for values of the cutoff greater than $4$ is extremely challenging if we do not use Monte Carlo. The degree of divergence of the same diagram with the EPRL spin foam model is entirely unknown. Any calculation with known techniques is too complicated. In Section~\ref{sec:VertexEPRL} we study it using Monte Carlo. Computing this amplitude is a stress test for the Monte Carlo framework and a novel result for studying EPRL spin foam infrared divergences.

\begin{figure}[H]
    \centering
    \includegraphics[width=0.75\textwidth]{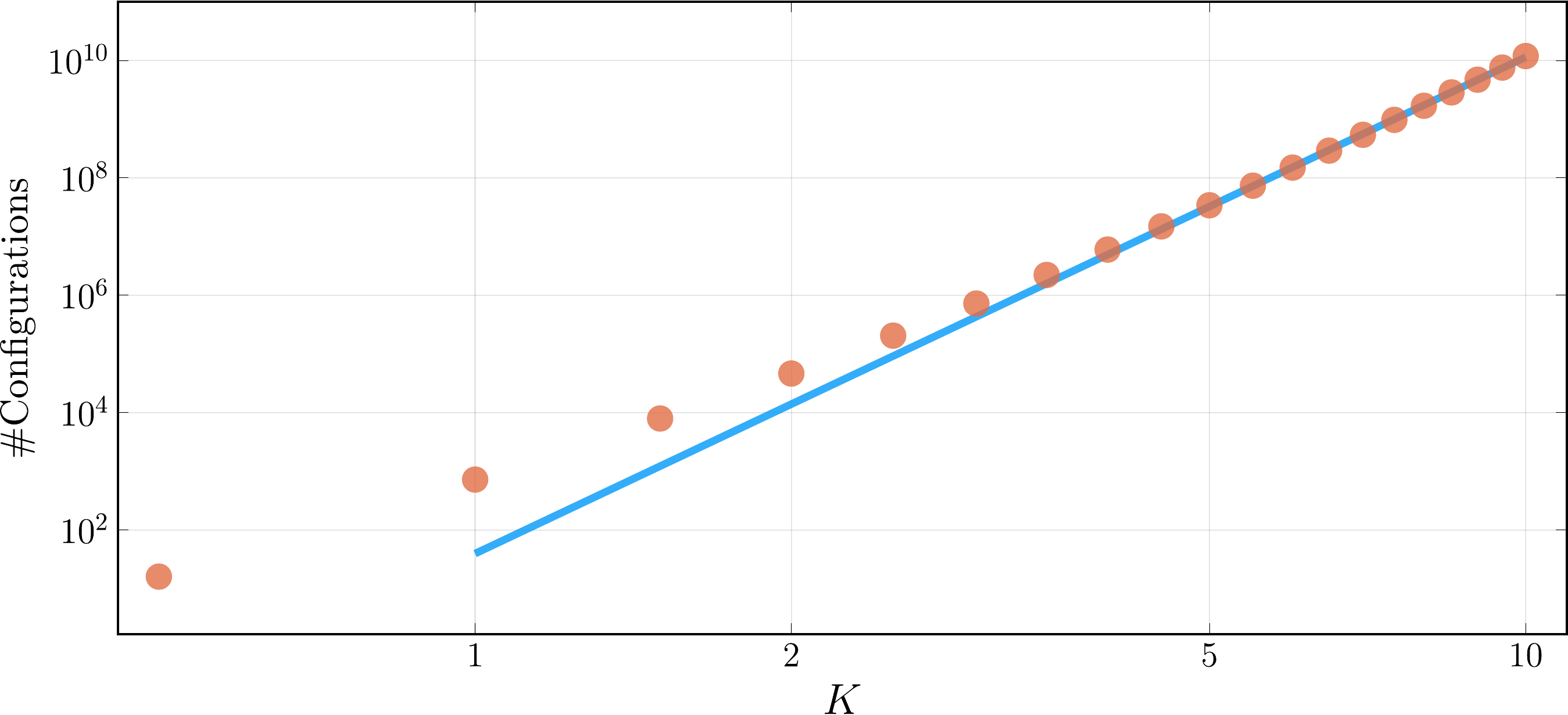}  
   \caption{\label{fig:numb_conf} Number of bulk spins configurations $j_f$ as a function of the cutoff $K$ of the vertex renormalization diagram.}
\end{figure}

In Figure \ref{fig:numb_conf}, we show the number of bulk spin configurations $j_f$ as a function of the cutoff $K$ for the vertex renormalization diagram. We only consider spin configurations that satisfy triangular inequalities. It is evident that the number of configurations to be summed increases as a power law with the cutoff $K$. Hence the convenience of using Monte Carlo. A simple numerical fit for $K \in \left[ 5, 10 \right]$ shows that the number of configurations qualitatively scales as $\sim 39.3 \cdot K^{8.5}$ for the vertex renormalization. For the self-energy diagram, in \cite{Frisoni:2021dlk}, the same fit for $K \in \left[ 5, 20 \right]$ showed that the number of configurations scales as $\sim 17.1 \cdot K^{5.6}$.

In all the amplitudes, to perform the numerical calculations with a modest amount of resources, we restrict the numerical calculation to the simplest non-trivial case of $j_b = \tfrac{1}{2}$ for both instances of boundary intertwiners $i_b=0$ and $i_b=1$. In the following sections, we will explicitly discuss the calculation's result only in the case of boundary intertwiner $i_b=0$. However, we performed the same analyses also with boundary intertwiners $i_b=1$. We find qualitatively identical results. Interested readers can find them in the detailed notebook in our public repository \cite{PaperRepo}. 

In the case of the EPRL model, we need to specify two more parameters to perform a numerical calculation. We fix the Immirzi parameter to $\gamma=0.1$. We choose this value to partially compare our results with the literature on the numerical evaluation of EPRL spin foam amplitudes \cite{Frisoni:2021uwx}. For similar reasons, we also choose the truncation parameter $\Delta l = 10$. This choice is also motivated by keeping the numerical task practical. The cost of resources increases rapidly with $\Delta l$, and literature \cite{Gozzini:2021kbt,Dona:2022dxs,Frisoni:2021uwx} suggests that for maximal spins of order $10$, the truncation $\Delta l =10$ is a good compromise between costs and precision. 

In the following sections, we use the same name for the amplitudes $A_{se}$ and $A_{vr}$ with both the BF $SU(2)$ and the EPRL model to keep the notation as clean as possible. The reader can uniquely identify which model the amplitude is computed with from the section. Finally, we use the term ``exact amplitude" referring to \eqref{eq:layered_amplitude} with a finite cutoff $K$ computed without resorting to Monte Carlo methods.

%%%%%%%%%%%%%%%%%%%%%%%%%%%%%%%%%%%%%%%%%%%%%%%%%%%%%%%%%%%%%%%%%%%%%%%%%%%%%%%%%%%%%%%%%%
\section{The melonic amplitude in the topological theory}
\label{sec:MelonBF}

In this section, we use the Monte Carlo framework described in Section~\ref{sec:sumMC} on the melonic self-energy transition amplitude in the topological $SU(2)$ BF model. This calculation aims to fine-tune and validate our choices of Monte Carlo parameters. 

\begin{figure}[H]
\centering
        \includegraphics[width=0.75\textwidth]{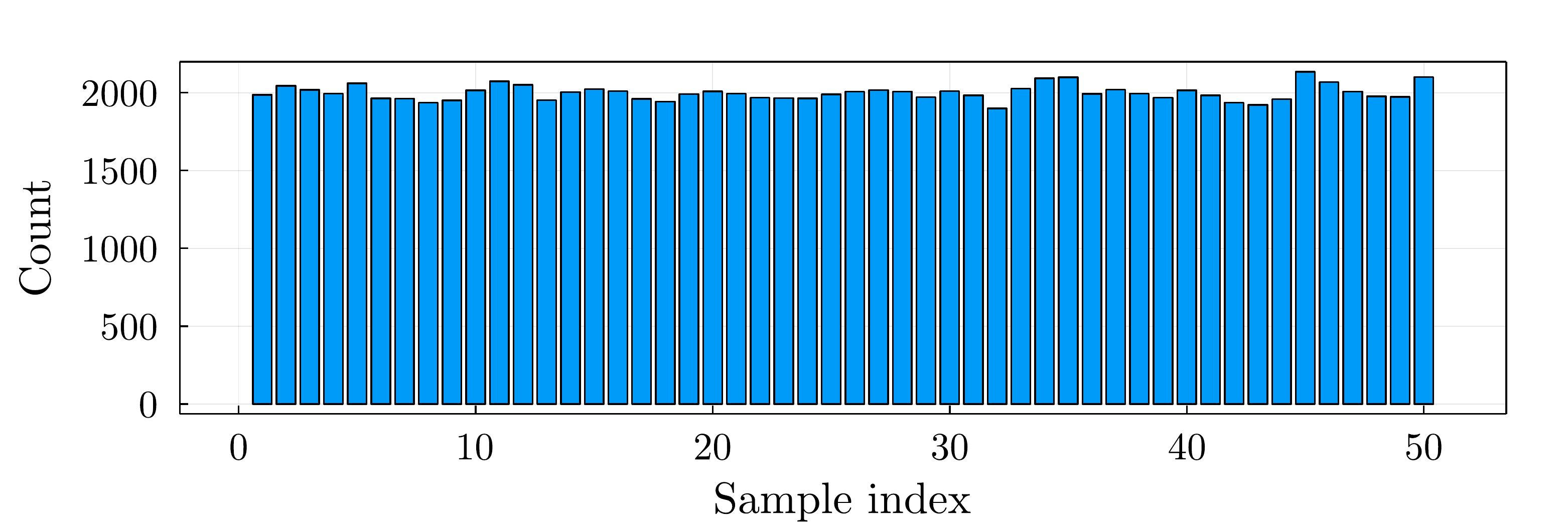}
\caption{\label{fig:sampler} Histogram of the sample with $100\,000$ configurations from the layer $k=10$ with $50$ bins. Each bin contains approximately $2\,000$ elements.}
\end{figure}

First, we test if the algorithm we use to sample the layers is equivalent to a uniform discrete probability distribution over the amplitude layers. We list \emph{all} the sets of bulk spins in the layer, and we map them in an interval of integers. Each element of the list is associated with its positions (we choose the order of the list arbitrarily but only once). We produce many samples using the Algorithm~\ref{algo:numericalsampling}.
We compute the samples' mean, variance, and skewness and check if they are compatible with the corresponding quantities of a discrete uniform probability distribution. We tested every layer of this amplitude and found excellent agreement. For brevity, we report the analysis with a sample of $100\,000$ configurations of the amplitude layer with $k=10$ that contain $V_{\mathcal{J}_k} = 549\,406$ possible configurations. We report them in Table~\ref{tab:comparison} and Figure~\ref{fig:sampler}. The probability distribution generated with Algorithm~\ref{algo:numericalsampling} is equivalent to a uniform discrete probability distribution.
\begin{table}[H]
    \centering
    \begin{tabular}{llll}
        \hline
        Quantity & Expected                  & Sample & Difference (\%)\\\hline\hline
        Mean     & $274\,703.5 $            & $274\,864.7$ & 0.05 \%\\
        Standard Deviation &  $1\,587\,371.04$ & $1\,585\,998.51$ & 0.09 \% \\
        Skewness &                   $0$                      & $-0.00545$  & - \\\hline
    \end{tabular}
    \caption{We compare the first three momenta of the sample with the corresponding exact quantities of a discrete uniform probability distribution. We find excellent agreement.\label{tab:comparison}}
\end{table}

\begin{figure}[H]
\centering
        \includegraphics[width=0.9\textwidth]{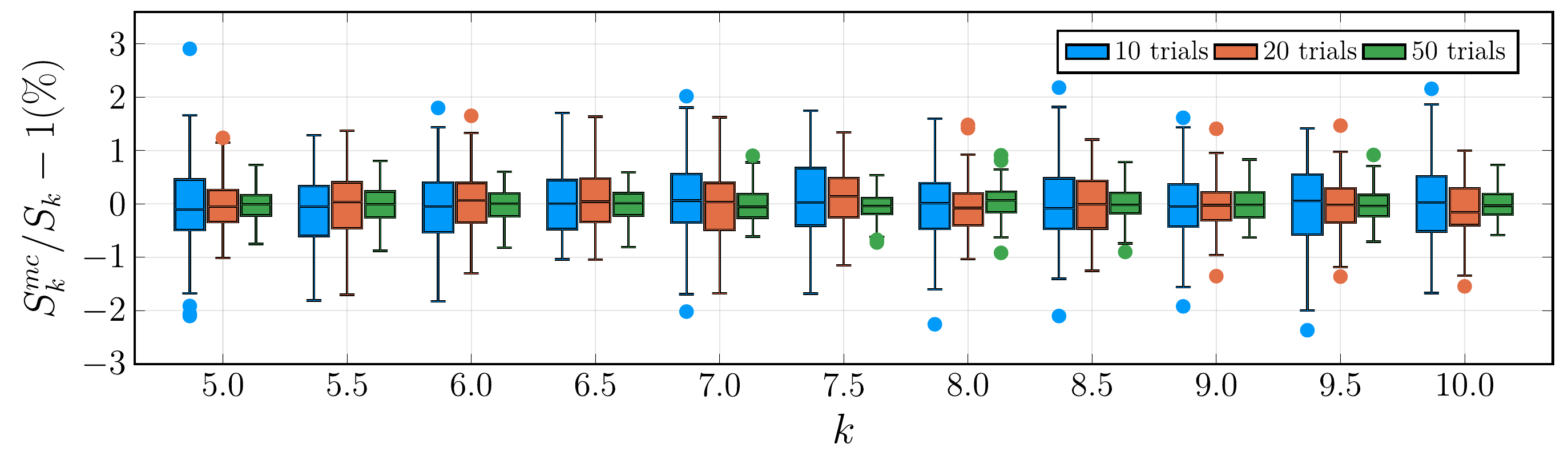}
\caption{\label{fig:error} Box plot of the average of the Monte Carlo evaluation of different amplitude layers. We consider the layers from $k=5$ to $10$. We repeat the estimate $100$ times. We look at $10$ trials (in blue), $20$ trials (in orange), and $50$ trials (in green). To ease the comparison, we plot the estimated value of the layer relative to the exact one.}
\end{figure}

We estimate the value and error of each amplitude layer with Monte Carlo repeating it $T=20$ times and computing the mean and standard deviation. We choose the number of trials after a simple test. We fix the size of the Monte Carlo sampling to $N_{mc}=1\,000$ to efficiently iterate and improve the analysis. We compute the average over $T=10$, $20$, and $50$ trials. We repeat it $100$ times to study the distribution of the average. The law of large numbers states that the distribution of the averages is normal with standard deviation given by the average standard deviation. We can visualize it using a box plot we report in Figure~\ref{fig:error}.
This qualitative analysis shows that the tails of the distribution with $10$ trials are very long. The standard deviation with just $10$ trials is not a reasonable estimate of the error of the Monte Carlo estimate of the amplitude's layer. This observation is independent of the layer. With $20$ trials, the first and third quartiles are reduced to half, making it a better option. A similar observation is valid for $50$ trials, albeit more resource intensive. We use $20$ trials as a good compromise between precision and simplicity. Nevertheless, the error between the Monte Carlo estimate relative to the exact quantity is always a few percent with $N_{mc}=1\,000$ and using $20$ trials instead of $10$ improve its estimate from $1\%$ to $2\%$. The gain in the error estimate using $50$ trials is marginal and does not justify the requirement of extra resources.

\medskip

\begin{figure}[H]
\centering
        \includegraphics[width=0.9\textwidth]{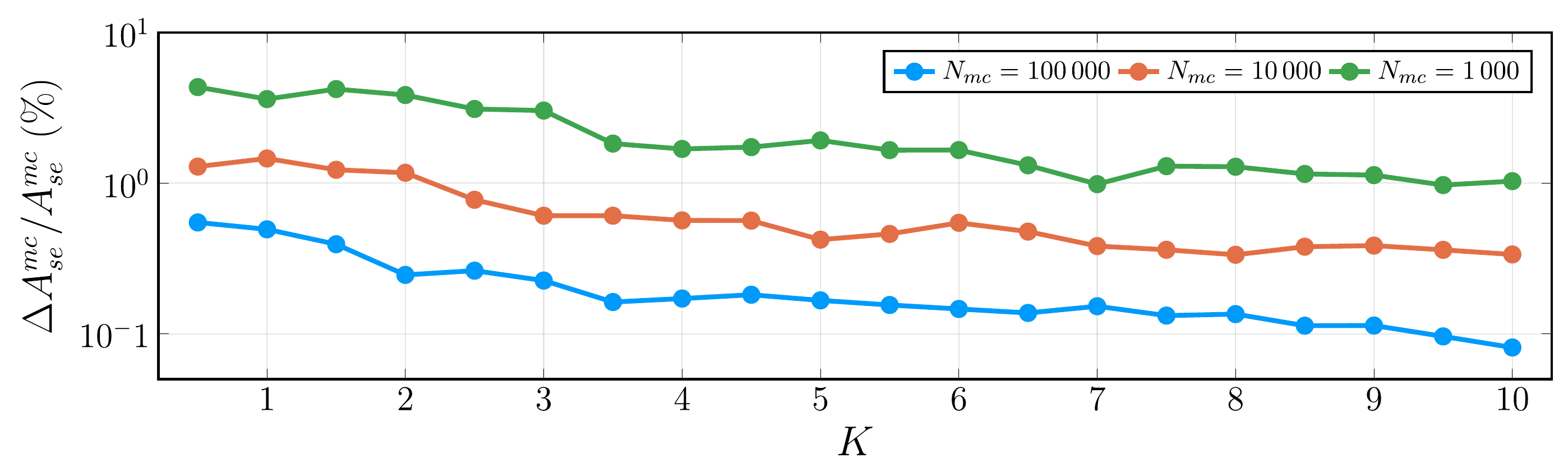}
\caption{\label{fig:SE_BF_relative_error} Relative error of the $SU(2)$ BF melonic amplitude as a function of the cutoff computed with $20$ trials. We compare different Monte Carlo sampling sizes $N_{mc}=1\,000$ (green), $N_{mc}=10\,000$ (orange), and  $N_{mc}=100\,000$ (blue) samples.}
\end{figure}

\begin{figure}[H]
\centering
        \includegraphics[width=0.9\textwidth]{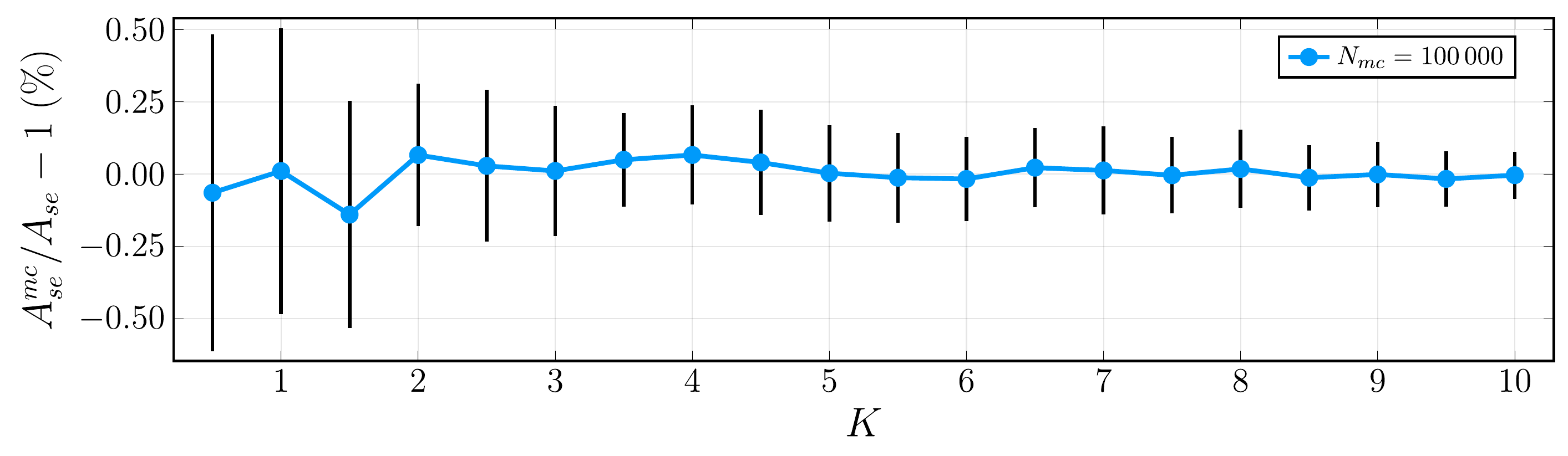}
\caption{\label{fig:SE_BF_error_bars} Monte Carlo estimates of the $SU(2)$ BF melonic amplitude relative to the exact value with $N_{mc}=100\,000$ and $T=20$ trials in each layer.}
\end{figure}

We study the Monte Carlo estimate of the amplitude as a function of the cutoff $K$ for three different sample size choices of $N_{mc}=1\,000$, $N_{mc}=10\,000$, and $N_{mc}=100\,000$. We average the calculation of $T=20$ trials for each layer and sum them to get the amplitude. We compute the error on the amplitude from the standard deviation of each layer. We compare the relative error on the amplitude as a function of the cutoff for different sizes of Monte Carlo samples (see Figure~\ref{fig:SE_BF_relative_error}).
For all three sample sizes, the relative error on the amplitude is smaller than $1\%$. As expected, the error decreases for larger values of $N_{mc}$. The relative error for $N_{mc}=100\,000$ is smaller than $0.1\%$. We decide to use $N_{mc}=100\,000$ for all the other calculations we present in this section. We also plot the estimated value of the amplitude with  $N_{mc}=100\,000$ relative to the exact value with the estimated errors. We plot in Figure~\ref{fig:SE_BF_error_bars} the Monte Carlo estimate of the amplitude in relation to the exact value, with error bars corresponding to the standard deviation. 
The exact value of the amplitude is compatible with the Monte Carlo estimate within the errors. At first sight, one could be confused by the trend of the errors decreasing with the cutoff. The observation that the relative error of the various layers is almost constant (as we can infer from Figure~\ref{fig:error}) can easily explain this. However, the contribution to the amplitude of the outer layers (with larger $k$) is bigger than the others. A quick back-of-the-envelope calculation shows that if we add two quantities a few orders of magnitudes apart but with the same relative error, the relative error on the sum is smaller than both.

\medskip

We conclude our exploration by estimating the degree of divergence of this amplitude. The analytic calculation shows that the amplitude diverges with the cutoff as $\propto K^9$ at the leading order. Can we determine it numerically? We answer with a proof of concept analysis we use to validate the technique before applying it to more complex amplitudes where the analytic answer is unknown. We are not satisfied with a qualitative result. We could easily eyeball a line on the logarithmic plot of the amplitude as a function of the cutoff. However, this approach is only helpful if we know the degree of divergence. We need to perform a fit to determine it numerically. We start with a model function with a polynomial form
\begin{equation}
\label{eq:model}
    A(K) = c_1 K^a + c_2 K^{a-1}  \ .
\end{equation}
We limit ourselves to the leading and subleading order terms. In general, the amplitude diverges as a polynomial with all the powers of the cutoff. Using it as a model to fit our data would undoubtedly lead to overfitting as we want to use a maximum cutoff of $K=10$. We could compute this amplitude for larger values of the cutoff. However, the EPRL model's amplitude is too computationally demanding, and we must impose a small cutoff of $K=10$. We use this limitation as an excuse to use a small cutoff with the topological model and anticipate some problems arising from this choice. We fit using only the last 10 data points available.

We perform a simple least squares fit using the \texttt{Julia} package \texttt{LsqFit}. We find the exponent $a  = 8.81 \pm 11.99$ and coefficients $c_1=4.83 \pm  240.38$, $c_2 =  41.12 \pm 66.95$. Examining the uncertainties of the parameters, we conclude that the fit is clearly unreliable. Moreover, even if the fit value for the exponent $a$ looks compatible with the exact value $a=9$, we could not affirm it without knowing it in advance.

What is happening? The covariance between the coefficients $c_1$ and $c_2$ is huge. The fit procedure with a cutoff of order $10$ cannot distinguish between the contributions from the leading and sub-leading orders (for example, if the exact coefficients are $c_1/c_2 \approx 10$). 

\begin{figure}[H]
\centering
        \includegraphics[width=0.9\textwidth]{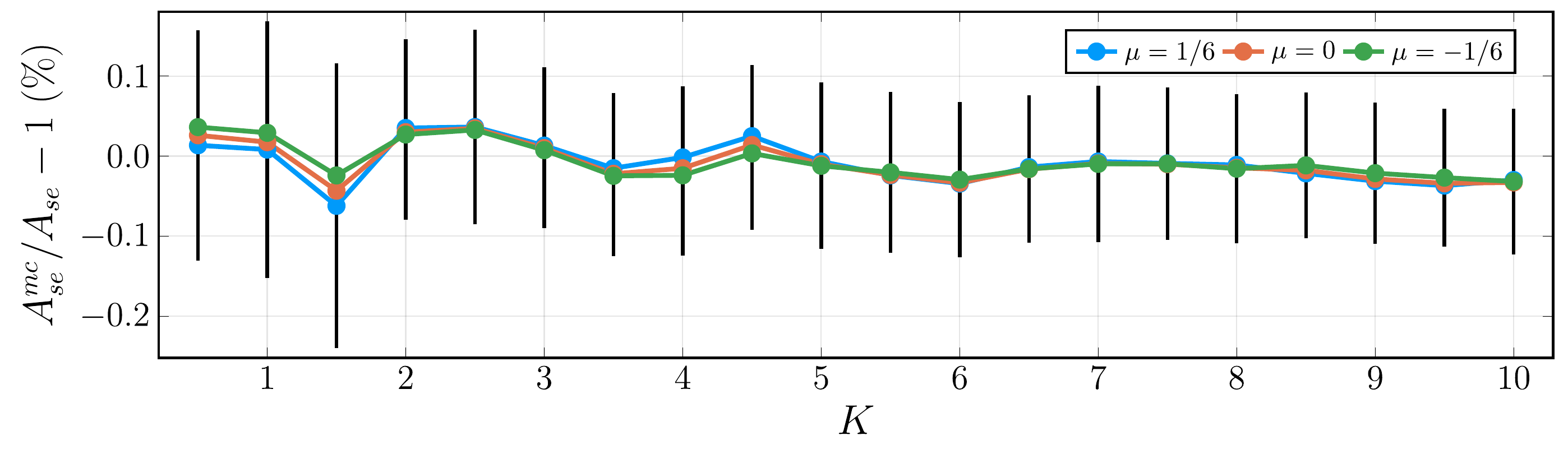}
\caption{\label{fig:SE_BF_error_bars_mu} Monte Carlo estimates of the $SU(2)$ BF melonic amplitude with $N_{mc}=100\,000$ and $T=20$ trials in each layer. We compare different values of the parameter $\mu$ in the face amplitude \eqref{eq:face_amplitude_mu}.}
\end{figure}

A simple solution to this impasse would be to extend the fit to larger values of $K$. In this way, the contribution of the leading order would dominate the sub-leading one. Since we are limited by a maximum cutoff $K=10$, we have to find a different solution. Alternatively, we can diminish the degree of divergence of the amplitude by changing the face amplitude introducing a tunable parameter $\mu$ 
\begin{equation}
\label{eq:face_amplitude_mu}
A_f(j_f) = (2j_f +1) \to (2j_f +1)^\mu    
\end{equation}
The case $\mu=1$ corresponds to the standard case, but if we set $\mu<1$ we lower the divergence of the amplitude. 
In general, the amplitude will diverge as $A_{se} \propto K^{6 \mu + p}$ where $p$ is a number we have to determine, and $6\mu$ is the contribution coming from the six unbounded sums over the bulk spins. 
We pretend we do not know that for $\mu=1$ the amplitude diverges as $A_{se} \propto K^{9}$ and therefore $p=3$. And we try to determine $p=a - 6 \mu$ fitting the amplitude with the same model 
\eqref{eq:model} for various values of $\mu=1/6$, $\mu=0$ and $\mu=-1/6$. We take this opportunity to check if the Monte Carlo estimate of the amplitude is as good as in the case $\mu=1$. We compare the Monte Carlo estimate of the amplitudes relative to their exact values with different $\mu$. All the relative error bars, computed over $T=20$ realizations as we did before, are within $0.1\%$, confirming that the Monte Carlo estimate is very accurate. We summarize the results in Figure~\ref{fig:SE_BF_error_bars_mu}.
We fit the amplitude with the model \eqref{eq:model}. The interpretation of the result, in this case, is more straightforward. We report the fitted coefficients in Table~\ref{tab:melon_fit_mu}.
\begin{table}[H]
    \centering
    \begin{tabular}{ccccc}
        \hline
                   & $a$             & $c_1$           & $c_2$ \\\hline\hline
        $\mu=1/6$  & $4.15 \pm 0.34$ & $0.91 \pm 1.24$ & $6.38 \pm 0.17$   \\
        $\mu=0$    & $2.97 \pm 0.04$ & $2.01 \pm 0.03$ & $3.50 \pm 0.05$  \\
        $\mu=-1/6$ & $1.94 \pm 0.02$ & $2.83 \pm 0.02$ & $0.39 \pm 0.05$ \\\hline
    \end{tabular}
    \caption{Values of the coefficients of the model \eqref{eq:model} obtained fitting the amplitude with different values of $\mu$.\label{tab:melon_fit_mu}}
\end{table}

A few comments are in order. First, all the fits indicate clearly that $p=a-6\mu=3$. Second, notice that we are not worried that $a$ is not always compatible with the nearest integer value. This is an artifact of using just the leading order and next to the leading order of the polynomial in \eqref{eq:model}. We determined the degree of divergence of the amplitude as $A_{se}\propto K^{6\mu +3}$.

%%%%%%%%%%%%%%%%%%%%%%%%%%%%%%%%%%%%%%%%%%%%%%%%%%%%%%%%%%%%%%%%%%%%%%%%%%%%%%%%%%%%%%%%%%

\section{The vertex renormalization amplitude in the topological theory}
\label{sec:VertexBF}
Exact numerical calculations of spin foam amplitudes with many bulk faces are accessible only for simple models, but become infeasible when the number of faces is too large. We showcase the problem by looking at the vertex renormalization or 5-1 Pachner move amplitude with the topological $SU(2)$ BF spin foam theory. The issue with this computation is not the time we need to compute each term of the sums over the spins of the ten internal faces but the sheer amount of terms of these sums. With a cutoff $K=10$ on the sums, we need to compute $30\,788\,382\,715$ terms in total, $11\,892\,969\,195$ of which belongs to last layer of the amplitude. They are almost six orders of magnitude more than in the melonic diagram case. This is where using the Monte Carlo framework to perform the sums is necessary.

Motivated by the analysis of the melonic diagram, we average the Monte Carlo calculation of each layer over $T=20$ trials. We sum the layers' average to obtain the amplitude value for a given cutoff. We compute the amplitude variance by summing each layer's variance. We consider the standard deviation as the error of the amplitude. We perform the calculation with three different choices of sample sizes $N_{mc}=1\,000$, $N_{mc}=10\,000$, and $N_{mc}=100\,000$.

\begin{figure}[H]
\centering
        \includegraphics[width=0.9\textwidth]{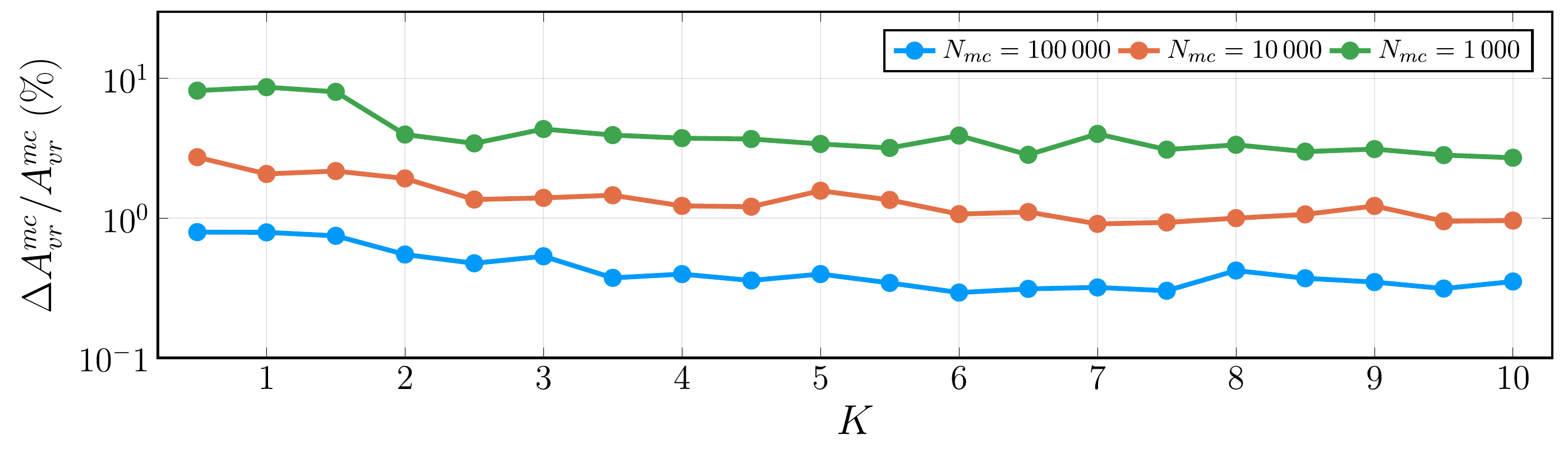}
\caption{\label{fig:VR_BF_relative_errors} Relative error on the Monte Carlo estimate of the $SU(2)$ BF vertex renormalization amplitude over $20$ trials. We compare different Monte Carlo sampling sizes $N_{mc}=1\,000$ (green),  $N_{mc}=10\,000$ (orange), and  $N_{mc}=100\,000$ (blue) samples. }
\end{figure}

In this case, we cannot compare with the exact value of the amplitude to evaluate the Monte Carlo technique. The exact value is not computable for cutoff $K=10$. It is exactly the reason we resort to Monte Carlo methods. We compare the relative error on the amplitude for the three sample size choices. We summarize the result of this analysis in the plot of Figure~\ref{fig:VR_BF_relative_errors}. 
We see that $N_{mc}=1\,000$ produces an estimate with a significant relative error between $8\%$ and $3\%$. For $N_{mc}=100\,000$ we find very modest relative errors between $0.8\%$ and $0.3\%$. The increase in the relative errors compared to the melonic diagram case is expected. Each layer of this amplitude contains a few orders of magnitude more elements than the corresponding layer in the melonic amplitude. 

We fit the amplitude computed with $N_{mc}=100\,000$ samples averaged over $T=20$ trials as a function of the cutoff $K$ using the model \eqref{eq:model}. We employ only the amplitude value as a function of the cutoff between $5$ and $10$ since we expect the matching of the functional form \eqref{eq:model} to be, at best asymptotic. The degree of divergence of this amplitude can be estimated analytically, finding $A_{vr} \propto K^{12}$ at the leading order. 

With a simple least squares fit, we find an unreliable result with an exponent $a  =  11.20 \pm 0.69$ and coefficients $c_1= 1.91 \pm  4.39$, and $c_2 = 1.70 \pm 10.82$. The situation is analogous to the case of the melon amplitude. To determine numerically the degree of divergence of this amplitude with a maximum cutoff of $K=10$ we modify the face amplitude as in \eqref{eq:face_amplitude_mu} such that the amplitude diverge as $A_{vr} \propto K^{10 \mu + p}$ with $p$ to determine. We want to keep the exponent $10 \mu + p$ as low as possible. Therefore, we pick three values of the weight $\mu=0$, $\mu=1/10$, and $\mu=1/5$. We find 
\begin{table}[H]
    \centering
    \begin{tabular}{ccccc}
        \hline
                   & $a$             & $c_1$           & $c_2$ \\\hline\hline
        $\mu=0$    & $1.96 \pm 0.02$ & $0.18 \pm 0.01$ & $-0.05 \pm 0.02$ \\
        $\mu=1/10$ & $2.89 \pm 0.02$ & $0.15 \pm 0.01$ & $0.07  \pm 0.02$ \\
        $\mu=1/5$  & $4.00 \pm 0.04$ & $0.08 \pm 0.01$ & $0.26  \pm 0.02$ \\ \hline
    \end{tabular}
    \caption{Values of the coefficients of the model \eqref{eq:model} obtained fitting the amplitude $A_{vr}$ with different values of $\mu$.\label{tab:melon_fit_mu_vertex}}
\end{table}
All three results are compatible with the analytic value of $p=2$ resulting in the amplitude diverging as $A_{vr}\propto K^{10\mu +2}$. 
%%%%%%%%%%%%%%%%%%%%%%%%%%%%%%%%%%%%%%%%%%%%%%%%%%%%%%%%%%%%%%%%%%%%%%%%%%%%%%%%%%%%%%%%%%

\section{The melonic amplitude in the EPRL theory}
\label{sec:MelonEPRL}

In Sections~\ref{sec:MelonBF}, we computed the melonic self-energy spin foam amplitude with the topological BF $SU(2)$ model using Monte Carlo. We obtained a remarkably accurate amplitude approximation, employing only a fraction of the computational resources. Is the Monte Carlo technique applicable to spin foam amplitudes with the EPRL model, and is it equally successful?

Before discussing the calculation details, we must disentangle two different overlapping approximations. One is due to the Monte Carlo sampling procedure, while the other is a consequence of working with a finite truncation parameter $\Delta l$. In calculating the melonic amplitude, we have access to the public data from \cite{Frisoni:2021uwx} that employs an extensive truncation $\Delta l =20$. We borrow that data to\ perform a detailed study of the truncation approximation independently from the Monte Carlo one. 

We mitigate the dependence from a specific choice of truncation $\Delta l$ using an extrapolation technique (see Appendix~\ref{app:Aitken}). This idea was first introduced in \cite{Frisoni:2021uwx}, further formalized in \cite{Dona:2022dxs}, and additionally utilized in \cite{Dona:2022vyh}. Our understanding of the extrapolation technique applied to EPRL spin foam amplitudes with finite truncation has improved considerably. In this section, we revisit it in a new light. We explicitly show that its principal hypothesis is satisfied and compare possible alternatives.

\begin{figure}[H]
\centering
        \includegraphics[width=0.9\textwidth]{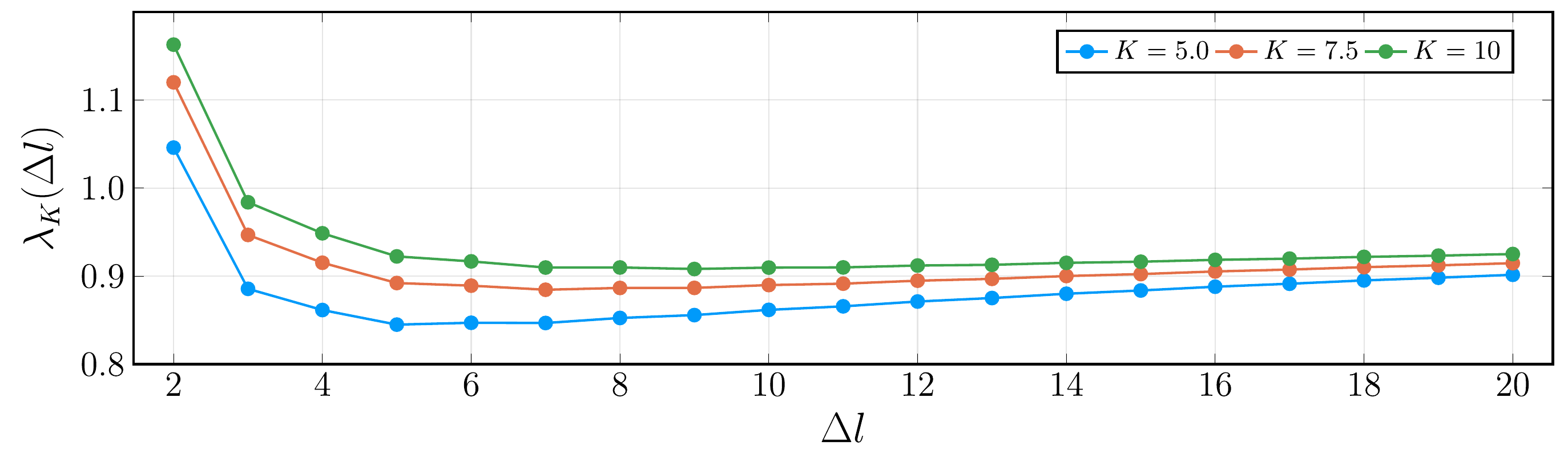}
\caption{\label{fig:lambdaK} Plot of the ratio \eqref{eq:ratio_amplitude} for various values of the cutoff $K=10$, $K=7.5$, $K=5$. A horizontal asymptote is evident even for modest values of the truncation.}
\end{figure}

The amplitude $A_{se}(K, \Delta l)$ at fixed cutoff $K$ is a sequence in the truncation parameter $\Delta l$. Since the EPRL vertex amplitude is well-defined, the limit of infinite truncation is finite, and we can approximate it using the Aitken delta squared method. 
\begin{equation}
    \label{eq:extrapolation}
    {A}^{(ex)}_{se}(K, \Delta l) = \frac{A_{se}(K, \Delta l)A_{se}(K, \Delta l-2) - A_{se}(K, \Delta l-1)^2}{A_{se}(K, \Delta l) -2 A_{se}(K, \Delta l-1)+A_{se}(K, \Delta l -2 )}
\end{equation}
The sequence ${A}^{(ex)}_{se}(K, \Delta l)$ converges to ${A}_{se}(K)$ faster than ${A}_{se}(K, \Delta l)$. Therefore approximating the limit with the truncation of the sequence ${A}_{se}(K)\approx {A}^{(ex)}_{se}(K) \equiv {A}^{(ex)}_{se}(K, \Delta l)$ is, in general, a better approximation than using the truncation of ${A}_{se}(K, \Delta l)$.
The extrapolation is effective if the rate of convergence of the amplitude is at least linear in $\Delta l$, as discussed in Appendix~\ref{app:Aitken}. For this purpose, we study the ratio 
\begin{equation}
\label{eq:ratio_amplitude}
    \lambda_{A_{se}(K)}(\Delta l) = \frac{A_{se}(K, \Delta l) - A_{se}(K, \Delta l-1)}{A_{se}(K, \Delta l-1) - A_{se}(K, \Delta l-2)} \ .
\end{equation}
If the limit of the ratio \eqref{eq:ratio_amplitude} for infinite truncation is smaller than $1$, the convergence of $A_{se}(K, \Delta l)$ is linear. Proving numerically the existence of the limit is challenging. Therefore, we settle with some numerical evidence for linear convergence. Moreover, truncating the sequence of extrapolations ${A}^{(ex)}_{se}(K, \Delta l)$ to a finite $\Delta l$ to approximate its limit is reliable only if the ratio approached (at least approximately) a horizontal asymptote.
Part of this analysis has already been performed in \cite{Frisoni:2021uwx}. Here we re-propose it in light of our improved understanding. For all half-integers $K \leq 10$, the ratio \eqref{eq:ratio_amplitude} approaches a horizontal asymptote smaller than $1$. This behavior is evident for any $8\leq \Delta l \leq 20$ and legitimizes the extrapolation of the amplitude \eqref{eq:extrapolation}. The ratio approaches the asymptote from below. Therefore, we expect the extrapolations from larger truncation to increase. We summarize the analysis for some cutoff values in Figure~\ref{fig:lambdaK}. 

\begin{figure}[H]
\centering
        \includegraphics[width=0.9\textwidth]{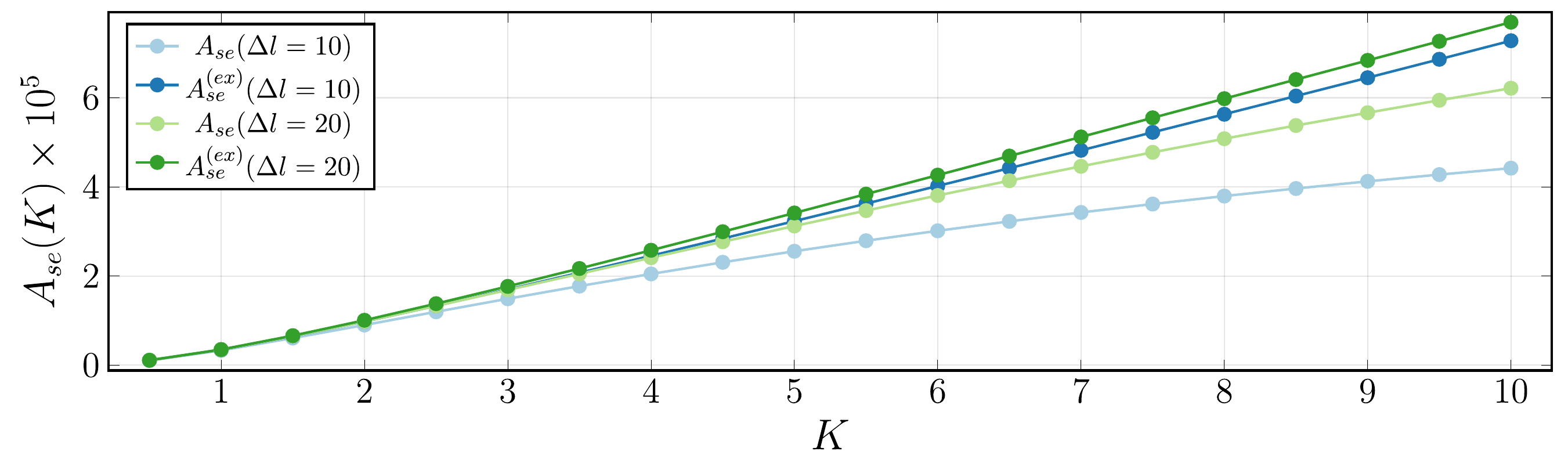}
\caption{\label{fig:extrapolationMelonEPRL} Comparison between the amplitude extrapolations and the truncated values ${A}_{se}(K,\Delta l)$ with $\Delta l=10$ and $\Delta l=20$ as a function of the cutoff. For larger cutoff values, the effect of truncation is critical. The extrapolation successfully mitigates this dependence.}
\end{figure}

How effective is the extrapolation \eqref{eq:extrapolation}? We answer this question by comparing the extrapolation obtained from $\Delta l =10$ and $\Delta l =20$. The last is usually inaccessible due to its exceptional computational cost, while the former is less precise but cheaper to compute. 
The value of the amplitude $A_{se}(K, \Delta l)$ changes substantially when we increase the truncation. For example, at cutoff $K=10$, the amplitude with $\Delta l =20$ is $40\%$ larger than the one with $\Delta l =10$. This was expected since the spins entering the calculation at $K=10$ are of order $20$, and the amplitude truncated at $\Delta l =10$ cannot approximate the real amplitude value well. However, we find a milder difference between the extrapolation \eqref{eq:extrapolation} with different truncations. With the same cutoff, the difference between the extrapolation done with truncation $\Delta l = 10$ and $\Delta l = 20$ is just $6\%$. In both cases, the extrapolations approximate the amplitude better than any truncated amplitude we have access to. We summarize the results in Figure~\ref{fig:extrapolationMelonEPRL}.

It is natural to question if extrapolating the amplitude sequence \eqref{eq:extrapolation} is the only way to proceed. We explored multiple alternatives (vertex-by-vertex, bulk amplitude, and layer-by-layer), and they all turned out to perform worst. We review in detail one of them. The amplitude is a sum limited to a maximum $K$ of contributions of layers \eqref{eq:layered_amplitude} we compute with a fixed truncation $\Delta l$
\begin{equation}
    \label{eq:selfenergylayer}
    A_{se}(K, \Delta l) = \sum_{k=0}^K S_k(\Delta l) \ .
\end{equation}
Each layer contribution $S_k(\Delta l)$ is a convergent sequence in the truncation parameter. We can use Aitken extrapolation on each layer contribution and obtain a sequence of amplitudes summing them
\begin{equation}
    \label{eq:extrapolation_layers}
     A_{se}^{(ex,S)}(K, \Delta l) = \sum_{k=0}^K S^{(ex)}_k(\Delta l) \qquad  \text{with} \qquad S^{(ex)}_k (\Delta l) = \frac{ S_k( \Delta l)S_k(\Delta l-2) - S_k(\Delta l-1)^2}{S_k(\Delta l) -2 S_k(\Delta l-1)+S_k(\Delta l -2 )} \ .
\end{equation}
We approximate the limit of the sequence of amplitude truncating the accelerated convergence sequence $A_{se}^{(ex,S)}(K) \approx A_{se}^{(ex,S)}(K, \Delta l)$ \eqref{eq:extrapolation_layers}.
Since the amplitude \eqref{eq:selfenergylayer} is a finite sum over layers, the sequence obtained extrapolating layer-by-layer \eqref{eq:extrapolation_layers} has the same limit as the sequence \eqref{eq:extrapolation} and both, of course, converge to the value of the amplitude $A_{se}(K)$ without any truncation. In practice, we do not have access to an arbitrarily large truncation but to a relatively small one, and we want to approximate the limit truncating the extrapolated sequences. Which one approximates the amplitude better is an open question that we can answer by exploring different possibilities.

\begin{figure}[H]
\centering
        \includegraphics[width=0.9\textwidth]{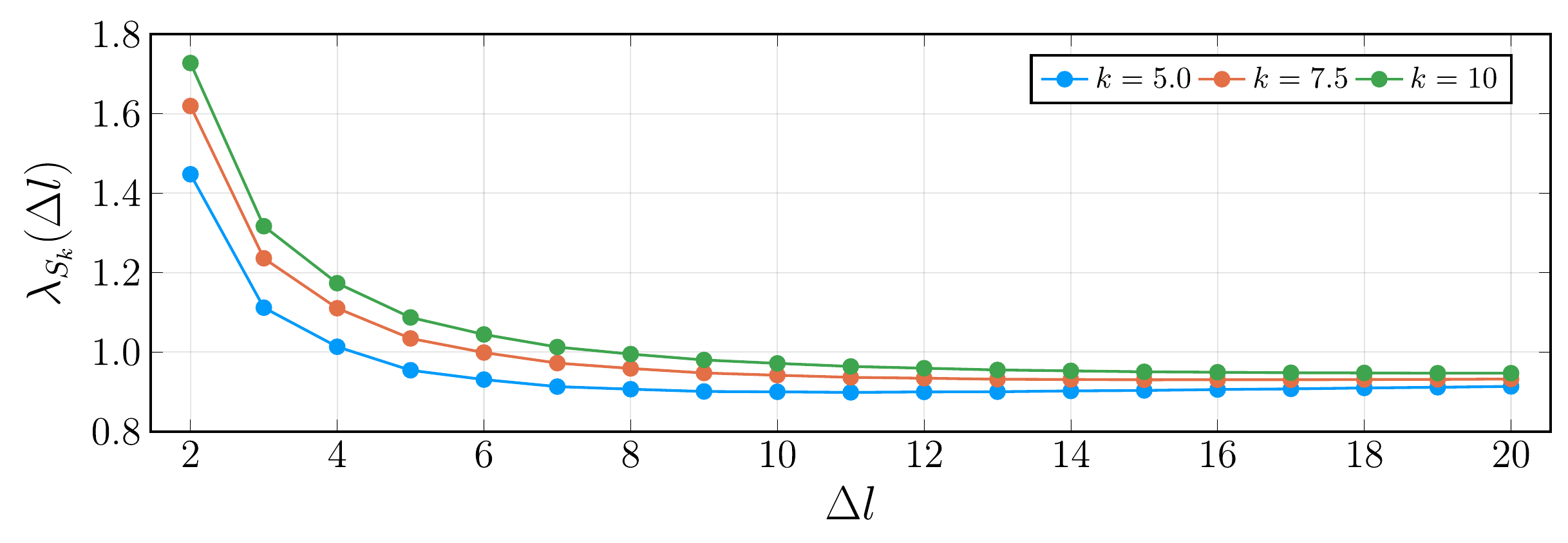}
\caption{\label{fig:lambdaKSk} Plot of the ratio \eqref{eq:ratio_layer} for various values $k=10$, $k=7.5$, $k=5$. A horizontal asymptote is evident but is reached for different truncation values.}
\end{figure}

\begin{figure}[H]
\centering
        \includegraphics[width=0.9\textwidth]{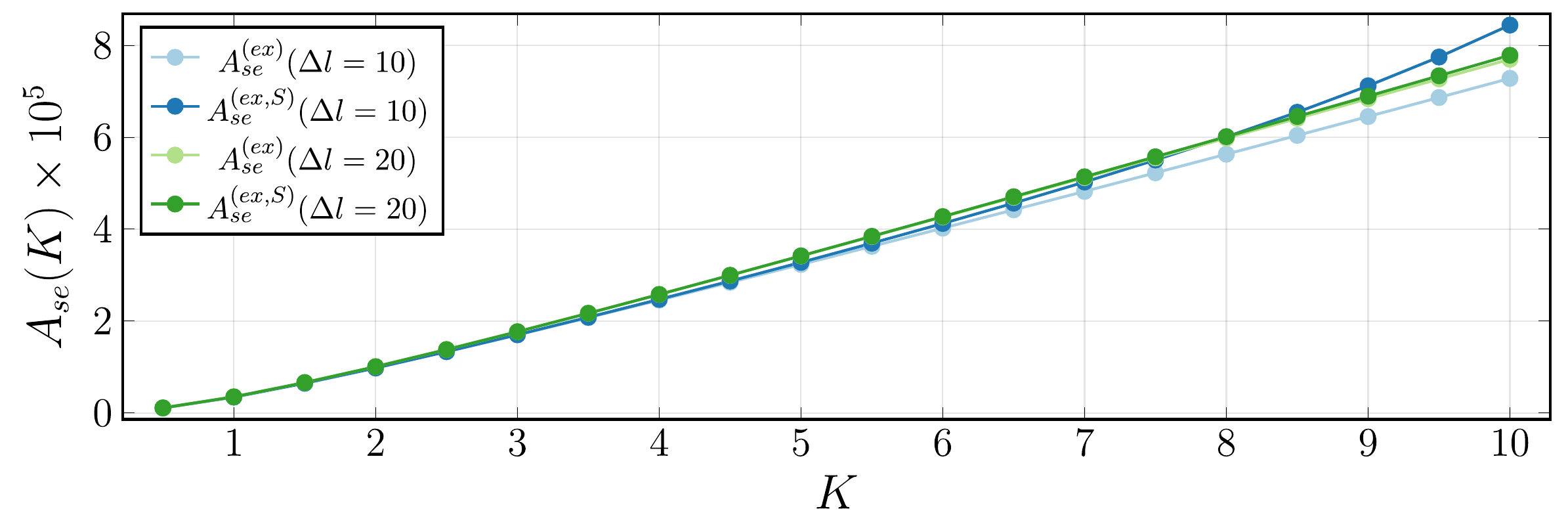}
\caption{\label{fig:extrapolationMelonEPRLlayer} Comparison of the melonic self-energy amplitude obtained extrapolating the amplitude layer-by-layer or the whole amplitude.}
\end{figure}

First, we verify that every layer amplitude is at least linearly convergent in the truncation parameter by studying the ratio
\begin{equation}
\label{eq:ratio_layer}
    \lambda_{S_k}(\Delta l) = \frac{S_k(\Delta l) - S_k(\Delta l-1)}{S_k( \Delta l-1) - S_k(\Delta l-2)} \ .
\end{equation}
We show some explicit examples in Figure~\ref{fig:lambdaKSk}. All the layer amplitudes are compatible with the linear convergence hypothesis. Differently from \eqref{eq:ratio_amplitude} the ratio \eqref{eq:ratio_layer} approach a horizontal asymptote from above. As a consequence, the extrapolation of the layer amplitudes decreases in value for increasing truncation. In particular, the last layer $k=10$ for $\Delta l = 10$ is still far from the horizontal asymptote. For this reason, we expect the extrapolation to change significantly if we increase the truncation from $\Delta l =10$ to, for example, $\Delta l =15$ or $20$ and approximate. This is not the case for lower levels $k\leq 5$. 
We extrapolate all the layer amplitudes and sum them. We truncate the sequence \eqref{eq:extrapolation_layers} to approximate its limit. It is useful to compare the result with the extrapolation of the whole amplitude \eqref{eq:extrapolation} using different truncation parameters. We summarize our findings in Figure~\ref{fig:extrapolationMelonEPRLlayer}.
For large cutoff, the extrapolation \eqref{eq:extrapolation_layers} with truncation $\Delta l = 10$ is much larger than its counterpart \eqref{eq:extrapolation}. At cutoff $K=10$, the layer-by-layer extrapolation \eqref{eq:extrapolation_layers} is $16\%$ larger than whole amplitude extrapolation \eqref{eq:extrapolation}. The difference reduces drastically if we repeat both extrapolations with a larger truncation $\Delta l = 20$. At cutoff $K=10$ the two differ only by $1\%$. In particular, the value of the extrapolation \eqref{eq:extrapolation_layers} decreases substantially. This agrees with what we observed studying the ratio \eqref{eq:ratio_layer}. With low truncation, the layer-by-layer extrapolation results in a poor amplitude approximation. In contrast, the two extrapolation schemes almost coincide for larger truncations.

In the following, we will study the Monte Carlo approximation of the amplitude and limit ourselves to a small truncation $\Delta l =10$ to conserve computational resources. We will use only the whole amplitude extrapolation scheme \eqref{eq:extrapolation} since it is the most accurate within this setting.

\begin{figure}[H]
\centering
        \includegraphics[width=0.9\textwidth]{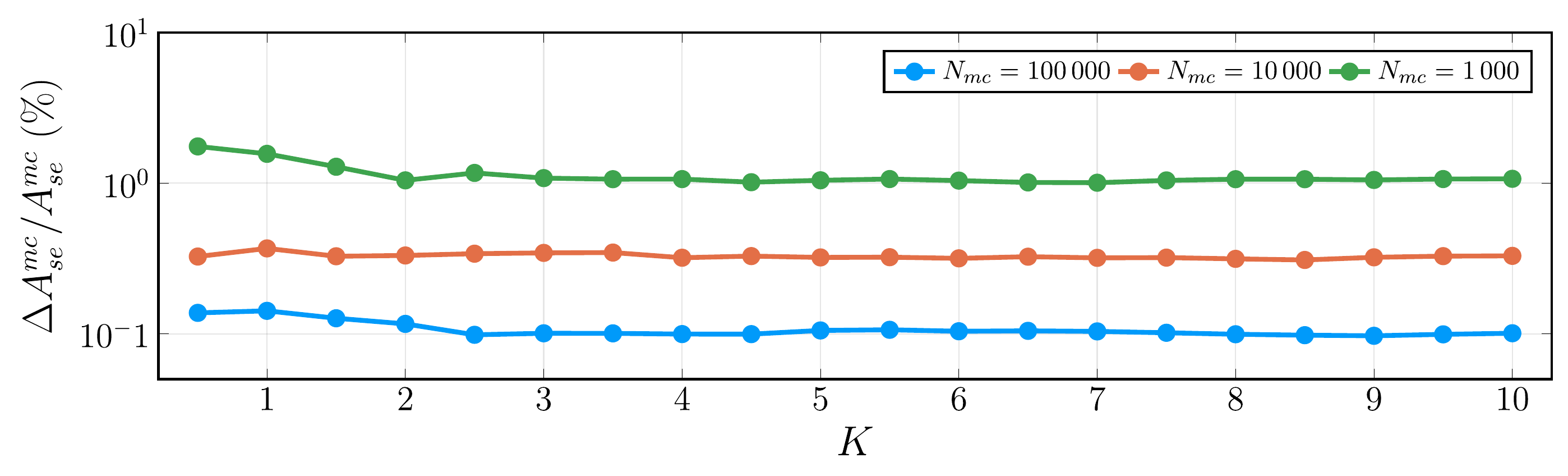}
\caption{\label{fig:SE_EPRL_relative_errors} Relative error of the EPRL melonic amplitude as a function of the cutoff computed with $20$ trials and truncation fixed to $\Delta l =10$. We compare different Monte Carlo sampling sizes $N_{mc}=1\,000$ (green), $N_{mc}=10\,000$ (orange), and  $N_{mc}=100\,000$ (blue) samples.}
\end{figure}

\begin{figure}[H]
\centering
        \includegraphics[width=0.9\textwidth]{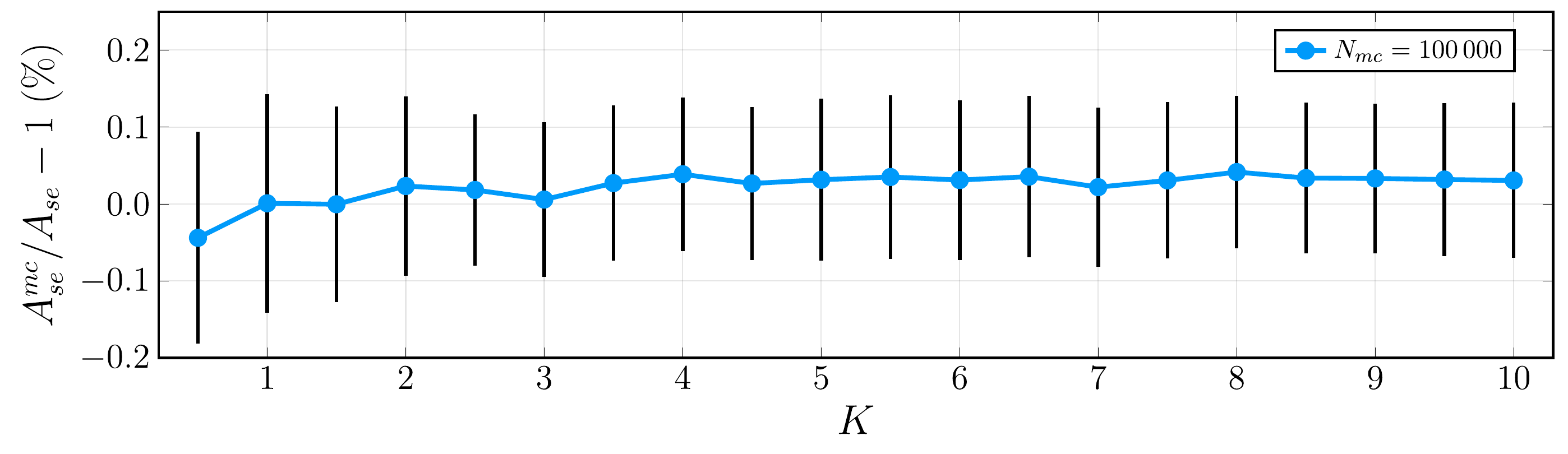}
\caption{\label{fig:SE_EPRL_error_bars} Monte Carlo estimate of the EPRL melonic amplitude with finite truncation $\Delta l =10$ relative to the exact value with $N_{mc}=100\,000$ and $T=20$ trials in each layer.}
\end{figure}

Next, we evaluate how effective Monte Carlo techniques are if applied to the bulk spin summations in the EPRL model using the melonic amplitude as a testing ground. Following the cost-benefit analysis of the first part of this section, we set the truncation to $\Delta l =10$. We use Monte Carlo to estimate the contribution to the amplitude of each layer averaging over $T = 20$ trials. The amplitude is given by the sum of the averages of the layers and its error by the square root of the total variance. 
We perform the calculation with three different choices of Monte Carlo sample sizes $N_{mc} = 1\,000$, $N_{mc} = 10\,000$, and $N_{mc} = 100\,000$. As displayed in Figure~\ref{fig:SE_EPRL_relative_errors} the relative error on the amplitude is more or less stable at $1\%$ for the small sample size and $0.1 \%$ for the large one.
The error is compatible with the analog error computed with the topological theory. We could have expected it as the distribution of the layers' amplitudes value is relatively flat. The dominant factor in the error is the ratio between the number of configurations in the layer and the Monte Carlo sample size, which is model-independent. 
At fixed truncation $\Delta l =10$, we can also compare the amplitude computed using Monte Carlo with the exact one. The amplitude computed using Monte Carlo is compatible with the exact value within the error. We summarize the comparison in the plot in Figure~\ref{fig:SE_EPRL_error_bars}.

How well the extrapolation technique \eqref{eq:extrapolation} is compatible with the Monte Carlo sum over the bulk spins? Instead of averaging over $T=20$ trials, we sum the layer amplitudes of each realization to obtain $20$ different realization of the amplitude at fixed truncation $\Delta l =10$. This is possible since each Monte Carlo estimate of each layer amplitude is independent. We extrapolate the amplitude \eqref{eq:extrapolation} for each trial and approximate it averaging over the trials and considering as error its standard deviation.  

\begin{figure}[H]
\centering
        \includegraphics[width=0.9\textwidth]{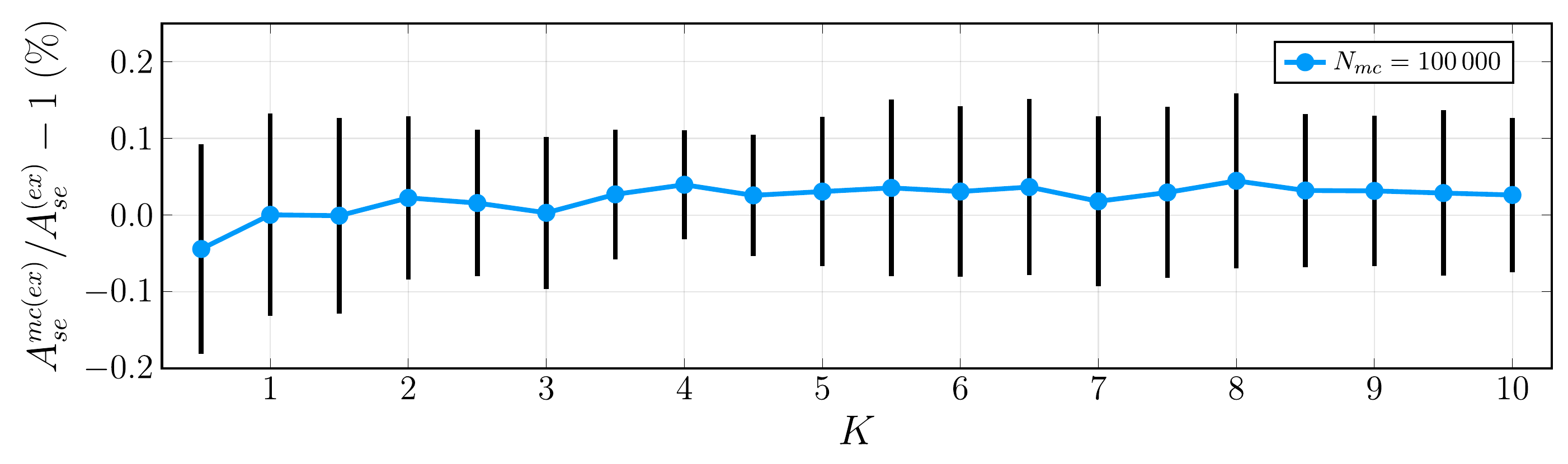}
\caption{\label{fig:SE_EPRL_error_bars_extrapolation}
Monte Carlo estimates of the EPRL melonic amplitude extrapolated from a finite truncation $\Delta l =10$ relative to the exact value with $N_{mc}=100\,000$ and $T=20$ trials in each layer.}
\end{figure}

We compare the extrapolation of the amplitude computed with Monte Carlo with the one calculated without that approximation. We find a $0.1 \%$ average error due to Monte Carlo for all cutoff values (see Figure~\ref{fig:SE_EPRL_error_bars_extrapolation}). This is compatible with the finite truncation case with $\Delta l =10$. However, this is just an error due to the Monte Carlo approximation. We expect it to be firmly subdominant with respect to the error due to the presence of the truncation despite the extrapolation. 

Finally, we determine the degree of divergence of the amplitude performing a numerical fit. The same calculation was already done in \cite{Frisoni:2021uwx} with an exact amplitude computation. We showed that using Monte Carlo allows us to estimate the amplitude with a $0.1\%$ error compared to the exact value (at fixed truncation). Therefore is no surprise that we find a good fit with 
\begin{equation}
    A(K) = c_1 K^a + c_2
\end{equation}
with $a=1.091 \pm 0.005$, $c_1=(6.186 \pm 0.093) \cdot 10^{-6}$, and $c_2=(-3.476 \pm 0.253) \cdot 10^{-6}$  that coincide with the result of \cite{Frisoni:2021uwx}.
We changed the model of the fit from the topological models, as \eqref{eq:model} gives unreliable results. If we generalize the face amplitude \eqref{eq:face_amplitude_mu} introducing a weight $\mu$, we conclude that the scaling of the amplitude is compatible with $A(K) \propto K^{6 \mu + p}$ with $p \approx -5$.

For completeness, we could look at the melonic divergence with different weight $\mu$ values as we did for the topological $SU(2)$ model. However, the result we obtain is unreliable. A more solid calculation requires a larger truncation and, consequently, way more computational resources that we currently do not have access to. We will comment on these issues in more detail in the next section. We leave this interesting consistency check to future work.

%%%%%%%%%%%%%%%%%%%%%%%%%%%%%%%%%%%%%%%%%%%%%%%%%%%%%%%%%%%%%%%%%%%%%%%%%%%%%%%%%%%%%%%%%%

\section{The vertex renormalization amplitude in the EPRL theory}
\label{sec:VertexEPRL}
\begin{figure}[H]
\centering
        \includegraphics[width=0.9\textwidth]{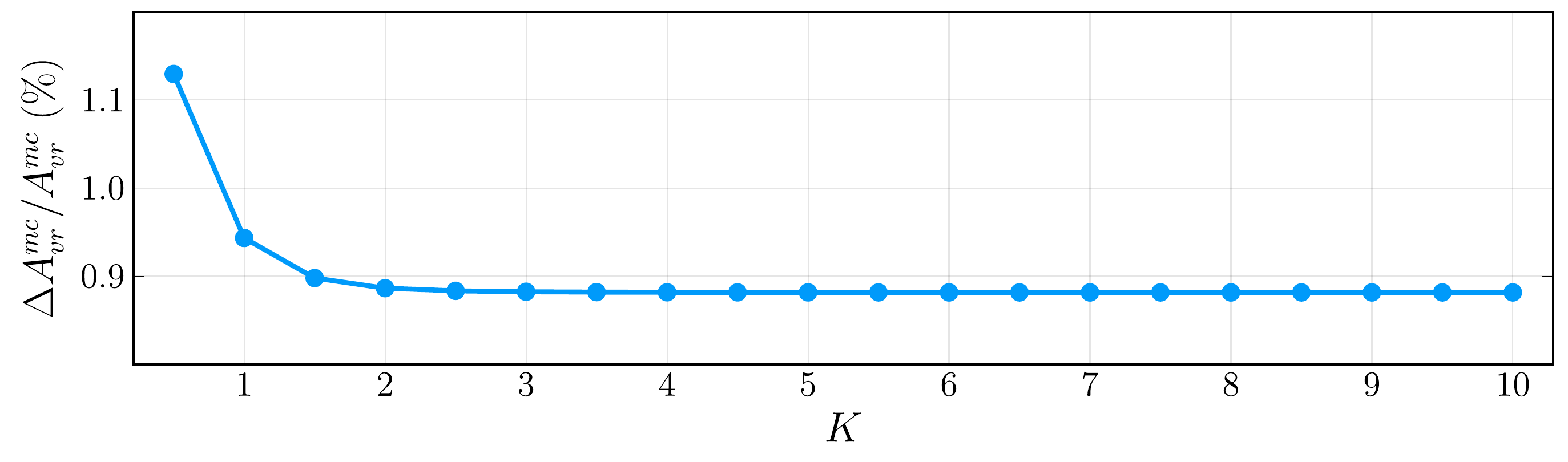}
\caption{\label{fig:VR_EPRL_error_bars} Relative error of the EPRL vertex renormalization amplitude as a function of the cutoff, computed with $20$ trials and truncation fixed to $\Delta l =10$. We use $N_{mc}=100\,000$ Monte Carlo samples.}
\end{figure}

Computing the degree of divergence of the EPRL vertex renormalization amplitude \eqref{eq:vertex} has never been attempted. The theory is too complex to do it numerically or analytically. We can use Monte Carlo to calculate this amplitude for the first time. We build upon the experience accumulated in the previous sections and our choices and approximations. For this reason, we use a truncation parameter $\Delta l=10$, and $N_{mc}=100\,000$ Monte Carlo samples. The calculation of this amplitude required $\sim 400$ CPU hours, which is a minimal fraction of what would be required without Monte Carlo. 
We compute the Monte Carlo error as the standard deviation of the amplitude over $T=20$ relative to its average. We find a very stable error of approximately $0.9\%$ for all values of the cutoff as summarized in Figure~\ref{fig:VR_EPRL_error_bars}
The relative error of the amplitude is curiously constant for increasing cutoff $K$. While this behavior could seem odd initially, it has a straightforward explanation. As we discuss below, the amplitude seems convergent. The amplitude layers decrease very fast. The contribution to the relative error of larger layers is irrelevant. 

The amplitude's statistical fluctuations are slightly bigger than in the EPRL self-energy amplitude. However, they are compatible with the errors of the topological model. It is not surprising since the ratio between the number of samples and the cardinality of the set we are summing over dominates the error of a Monte Carlo calculation.

\begin{figure}[H]
\centering
        \includegraphics[width=0.9\textwidth]{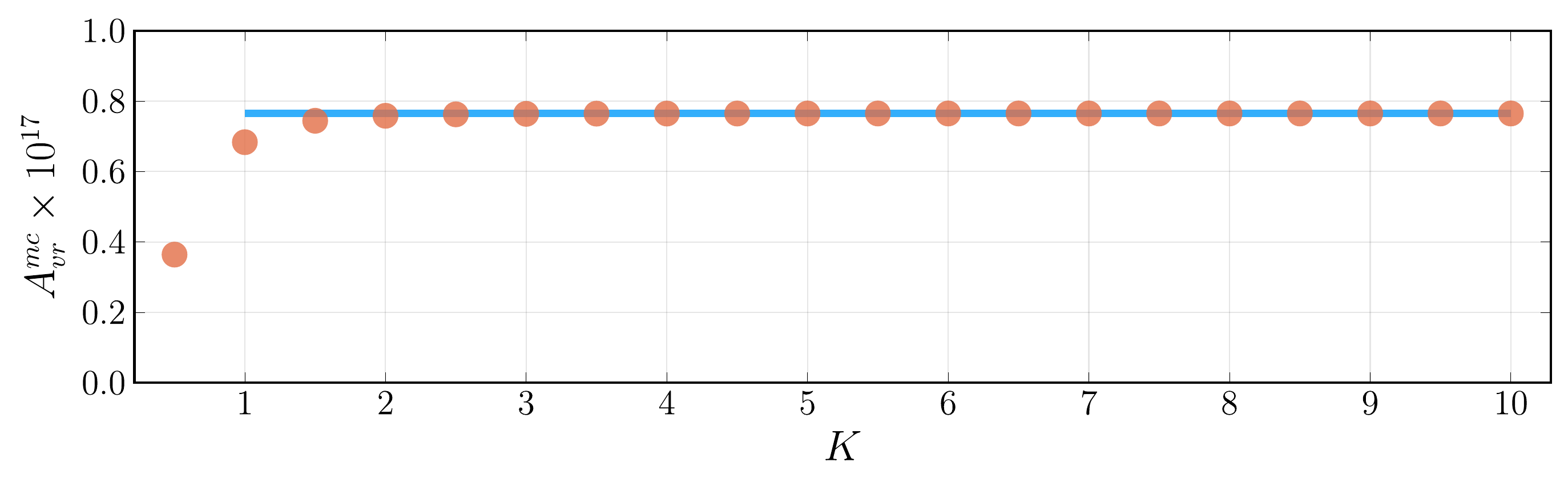}
\caption{\label{fig:VR_EPRL_fit} Monte Carlo  estimate of the EPRL vertex renormalization amplitude as a function of the cutoff $K$. We use $\Delta l = 10$, $N_{mc} = 100\,000$ and $T=20$ trials. We plot the extrapolated amplitude and the fit using the model \eqref{eq:VR_EPRL_fit_model}.}
\end{figure}

We reduce the dependence of the amplitude from the truncation using the extrapolation \eqref{eq:extrapolation} as discussed in Section~\ref{sec:MelonEPRL}. We show the value of the extrapolated amplitude as a function of the cutoff in Figure \ref{fig:VR_EPRL_fit}. The amplitude is essentially constant for cutoff $K > 2$. A power law fit is inadequate to capture the functional scaling of the amplitude. Therefore we opt for a model capturing the constant behavior plus a correction.
\begin{equation}
\label{eq:VR_EPRL_fit_model}
    A(K) = c_1 + c_2 / K \ .
\end{equation}
Fitting the amplitude result in $c_1 = 0.765 \pm 2.667 \times 10^{-5}$ and $c_2 = -0.0006 \pm 0.0002$. We should take these values with a grain of salt as they depend strongly on the model we decide to use. 
We are tempted to enhance the divergence of the amplitude by modifying the face amplitude \eqref{eq:face_amplitude_mu} introducing the weight $\mu$. We observe that increasing $\mu$ we need to increase $\Delta l$. Otherwise, the extrapolation technique fails in estimating the amplitude well.

\begin{figure}[H]
\centering
        \includegraphics[width=0.9\textwidth]{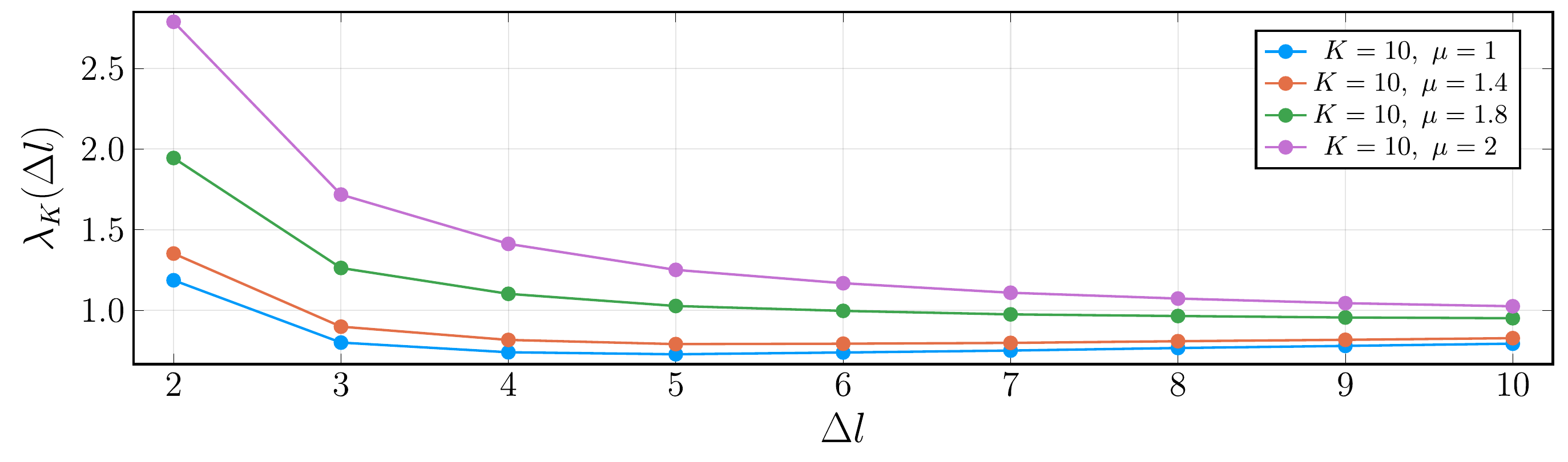}
\caption{\label{fig:VR_EPRL_lambda} Plot of the ratio \eqref{eq:ratio_amplitude} for increasing values of $\mu$ at cutoff $K=10$. For $\mu=1.8$ and $\mu=2$ the ratio has not reached a horizontal asymptote smaller than $1$ at $\Delta l=10$. Therefore the extrapolation \eqref{eq:extrapolation} cannot be applied.}
\end{figure}

We tried different face amplitudes weights and studied the amplitude ratios \eqref{eq:ratio_amplitude}. We show it in Figure~\ref{fig:VR_EPRL_lambda}. While for standard face amplitude $\mu=1$ and $\mu=1.4$, the ratio reaches a horizontal asymptote smaller than $1$ very soon, it is not the case for $\mu=1.8$, and $\mu=2$. We can see how for these two cases, the ratio is still decreasing and greater than $1$ at truncation $\Delta l =10$. Therefore, to obtain a reliable extrapolation we would need a larger truncation, not accessible with the computational resources at our disposal. 

\begin{figure}[H]
    \centering
        \begin{subfigure}[b]{0.45\textwidth}
        \includegraphics[width=1\textwidth]{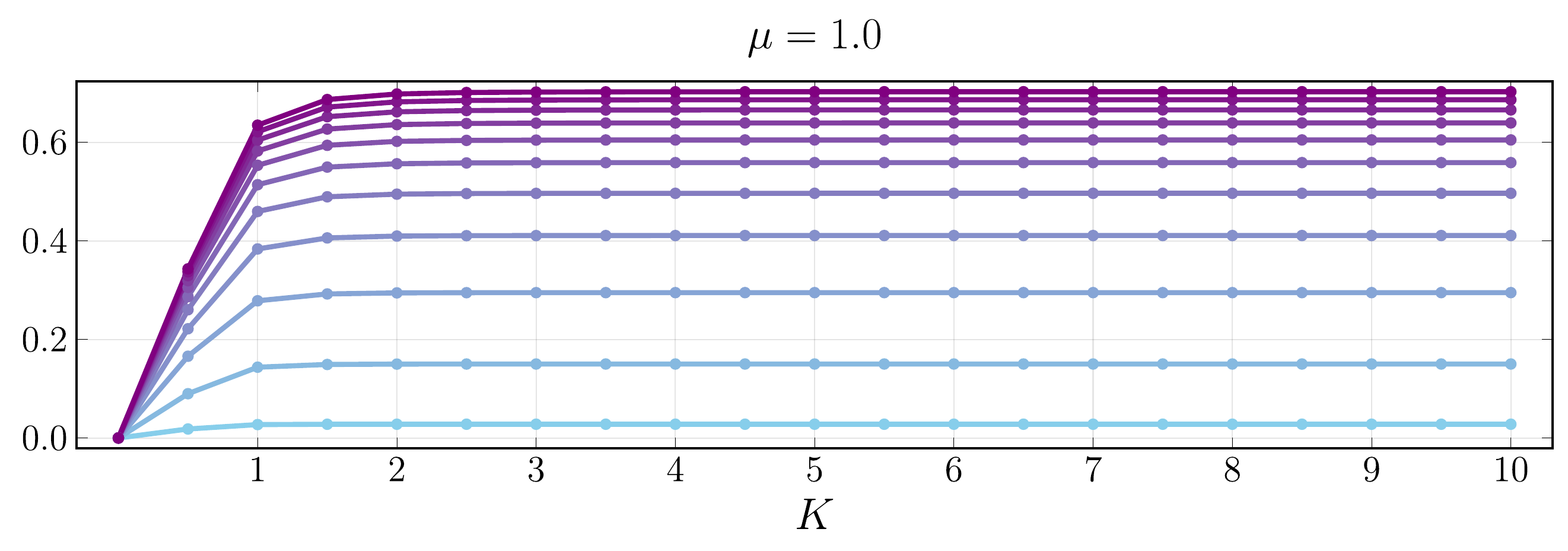}
     \end{subfigure}
    ~~~
        \begin{subfigure}[b]{0.45\textwidth}
        \includegraphics[width=1\textwidth]{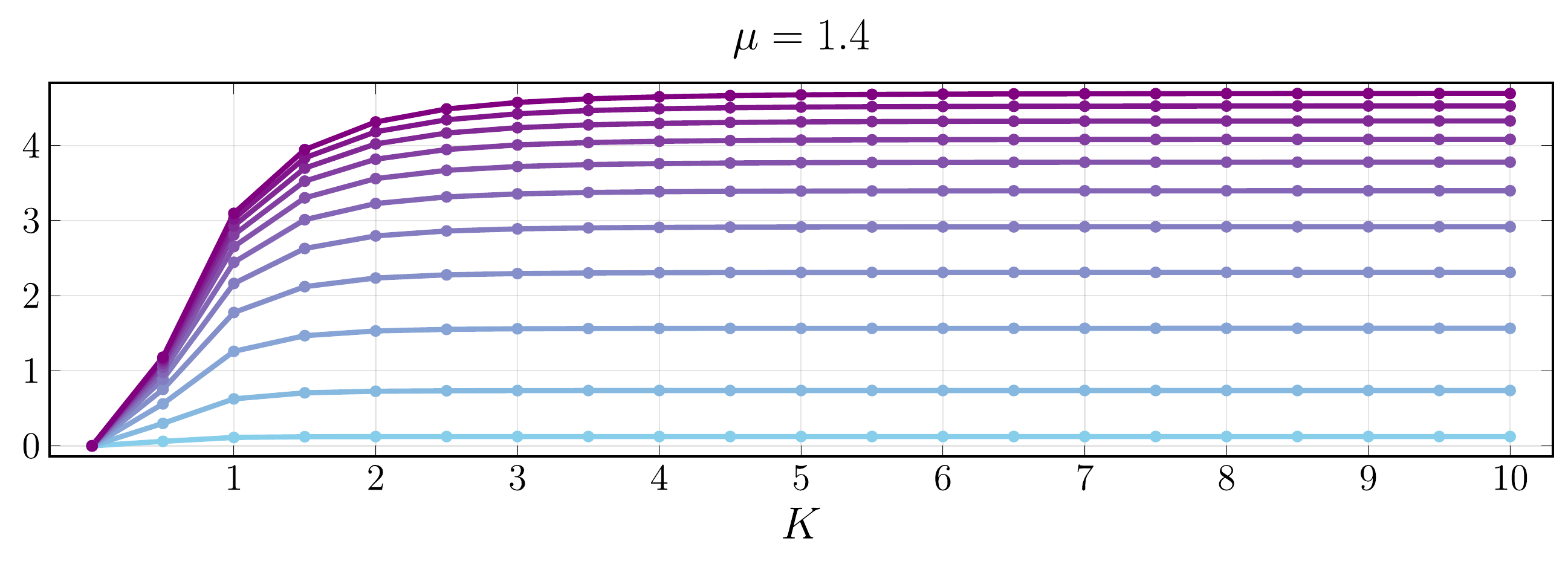}
    \end{subfigure} 
     ~~~
        \begin{subfigure}[b]{0.45\textwidth}
        \includegraphics[width=1\textwidth]{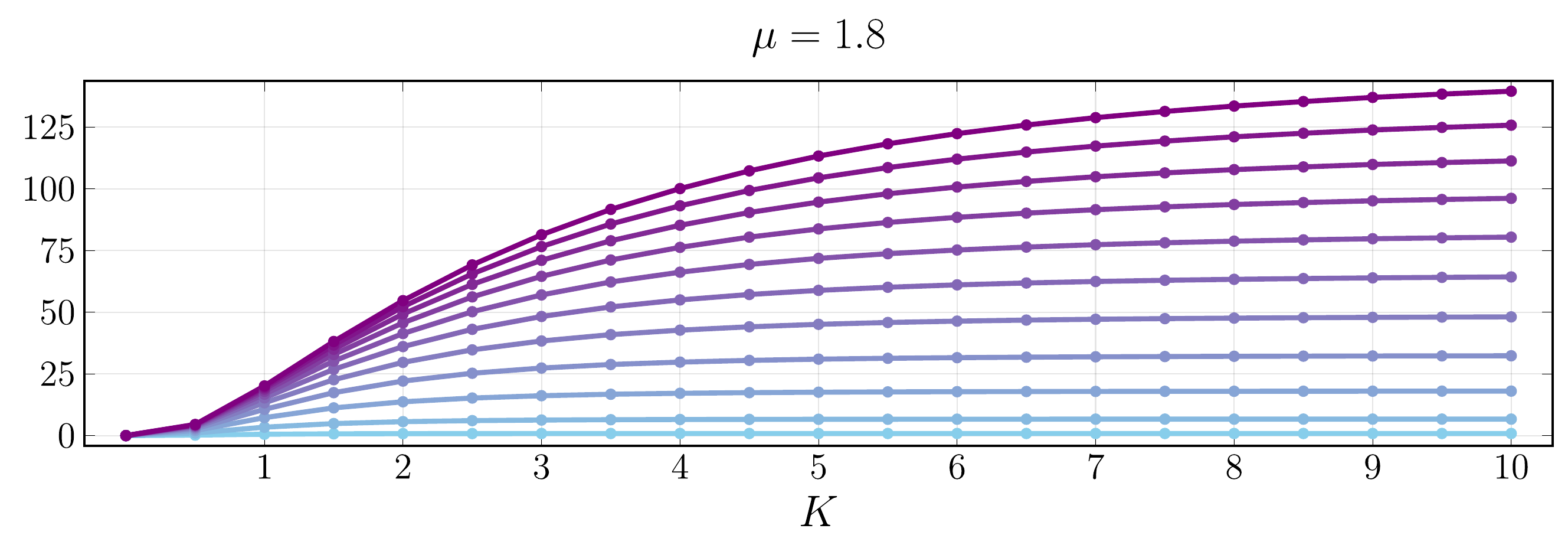}
    \end{subfigure}   
      ~~~
        \begin{subfigure}[b]{0.45\textwidth}
        \includegraphics[width=1\textwidth]{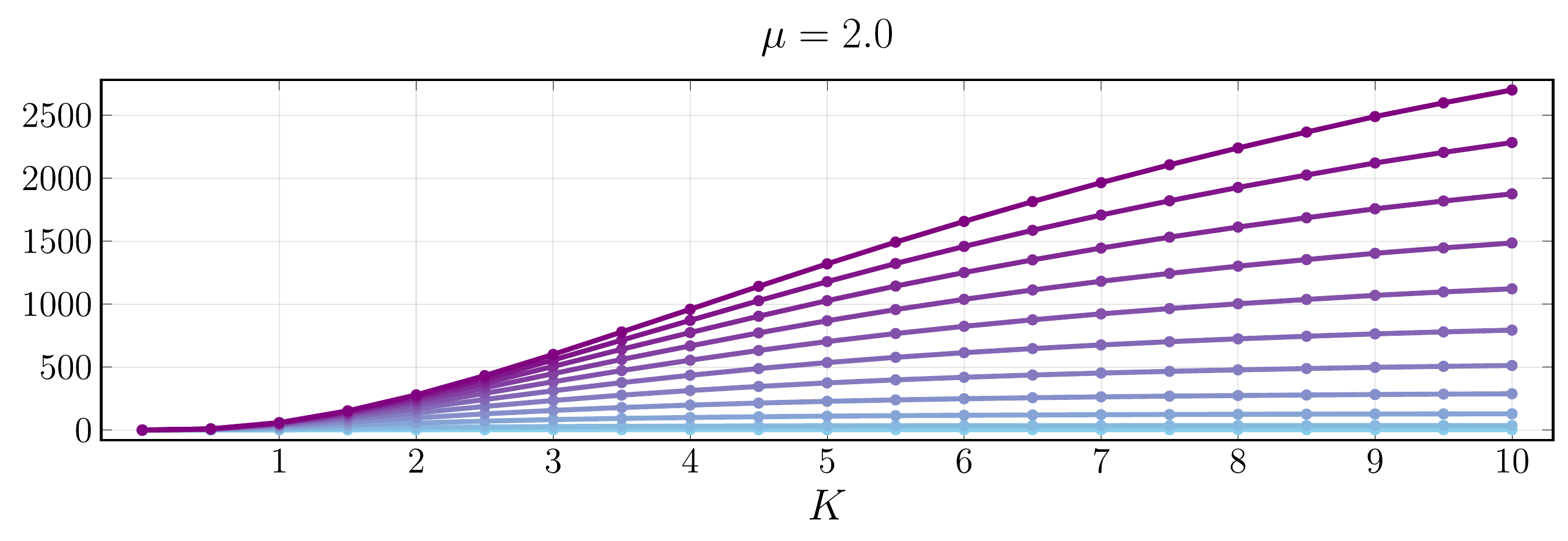}
    \end{subfigure}     
   \caption{\label{fig:VR_EPRL_increasing_mu} Plots of vertex renormalization EPRL amplitude $A_{vr}^{mc} \times 10^{17}$ for increasing values of the weight factor $\mu$. In each panel, we report all the curves obtained for increasing values of the truncation parameter $\Delta l$. The bottom curve (azure) corresponds to $\Delta l = 0$ while $\Delta l=10$ is the top one (purple).}
\end{figure}

We conclude that for this transition amplitude, the extrapolation scheme \eqref{eq:extrapolation} is sensible to the weight on the face amplitudes. The slower convergence of the amplitude sequence is also evident from the plot of the amplitude for different truncations as well displayed in Figure~\ref{fig:VR_EPRL_increasing_mu}.
For completeness, we report the fit of the convergent amplitude with $\mu = 1.4$ with the model \eqref{eq:VR_EPRL_fit_model}. We find $c_1=5.535 \pm 0.011$ and $c_2=-0.464 \pm 0.074$. Due to the invalidity of the extrapolation, we cannot perform a fit in the other two cases. Consequently, we cannot estimate $p$ in the divergence of the amplitude $A_{vr} \propto K^{10 \mu + p}$. Nevertheless, our numerical analysis shows strong indications that the infrared bubble of the EPRL theory is convergent.

%%%%%%%%%%%%%%%%%%%%%%%%%%%%%%%%%%%%%%%%%%%%%%%%%%%%%%%%%%%%%%%%%%%%%%%%%%%%%%%%%%%%%%%%%%

\section{Conclusion and Discussion}
\label{sec:conclusion}
The \texttt{sl2cfoam-next} library allows fast and reliable calculations of EPRL spin foam transition amplitudes. While it is optimized to compute vertex amplitudes, calculating a spin foam amplitude with many vertices and internal faces still presents a huge technical obstacle. To sum over the bulk degrees of freedom, we have to compute an enormous number of components that scale exponentially with the number of internal faces.

\medskip

We apply Monte Carlo to the spin foam bulk summations and show that it is a very promising strategy to overcome this obstacle. The complexity of the calculation depends on the number of Monte Carlo samples $N_{mc}$ we can freely choose. Of course, the result's precision depends on $N_{mc}$ and how we choose the probability distribution of the Monte Carlo sampling. We decide to use a uniform probability distribution. We acknowledge it is not the optimal choice as it equally weights all the bulk spins configurations. However, it is efficient and allows us to parallelize the sampling algorithm for a single amplitude across multiple threads.
Moreover, we can use it democratically with any amplitude. Alternatively, we could abandon uniform sampling in favor of Markov Chain Monte Carlo (MCMC) methods. This would result in faster convergence, but on the other hand, the sampling algorithm for calculating a single amplitude would no longer be parallelizable. We leave the study and implementation of MCMC to bulk degrees of freedom for future works. Finally, in this paper, we used intertwiners as boundary states. In other cases, one could attempt to take advantage of the state's properties to perform importance sampling Monte Carlo. For example, in the case of extrinsic boundary states, it would make sense to tailor the probability distribution sampling from a normal distribution. One has to deal with the well-known sign problem with highly oscillatory distributions. We leave the analysis of different boundary states for future works.

\medskip

We evaluate the proposed strategy and discuss its choices for computing the melonic self-energy and the vertex renormalization spin foam amplitudes, with the $SU(2)$ BF and the EPRL theory. These are amplitudes with many vertices and internal faces, providing a good test. In particular, the topological theory is convenient as partial analytical calculations are possible and help us evaluate the performance of our method. The calculations of the amplitudes with Monte Carlo are surprisingly effective already with a modest number of Monte Carlo samples $N_{mc}=100\,000$. To appreciate this result remember that the vertex renormalization amplitude contains $\approx 10^{10}$ possible (non trivially vanishing) amplitudes. Obtaining a good approximation computing just order $\approx 10^5$ of them is remarkable. 
We find small uncertainties of $0.1\%$ and $0.9\%$ for the EPRL self-energy and the vertex renormalization amplitude, respectively. Similar results hold for the $SU(2)$ BF model suggesting that the dominant factor in the error is the ratio between the cardinality of the space of all the possible spin configurations and the (square root of the) Monte Carlo samples. The calculations in the EPRL theory are carried out with a finite truncation $\Delta l =10$. We alleviate the dependence of the result using extrapolation techniques to accelerate the convergence of the amplitude. We explore different extrapolation schemes. Extrapolating the full amplitude at a finite cutoff is the most convenient option. We also formalize the regime of validity of the extrapolation scheme and develop a test to verify if the amplitude falls into it. 

\medskip

At the same time, analyzing the divergence of these amplitudes for the EPRL theory is an essential step toward understanding the theory's continuum limit. While the self-energy amplitude is already studied in the literature \cite{Frisoni:2021uwx,Dona:2018pxq,Riello:2013bzw}, for the vertex renormalization amplitude, we know only a loose upper bound \cite{Dona:2018pxq}. The amplitude was too complex to try any numerical or analytical calculation. The estimate we provide in this paper is a complete and important novelty. Performing the sums over the bulk degrees of freedom with statistical methods is enough to confirm the linear divergence of the melonic self-energy amplitude. 

The numerical evaluation of the EPRL vertex renormalization spin foam amplitude provides a convincing argument to claim its convergence. It is a shocking result as it contradicts any intuition we could get from the analytical calculations of the topological models where the amplitude is more divergent than the melonic self-energy one. 

Our result is a numerical computation of the amplitude, not analytical proof. One should always keep in mind its limitations. We performed a calculation with fixed boundary spins $j_b=\tfrac{1}{2}$, Immirzi parameter $\gamma = 0.1$, an extrapolation based on the truncation $\Delta l = 10$, and a uniform cutoff on all the faces limited to $K\leq 10$. Technical limitations and convenience dictate some choices. Nevertheless, we explored alternatives when possible, and the result seems general. 
Numerically we infer that the convergence of the vertex renormalization amplitude is determined by the destructive interference of the vertex amplitudes' oscillations of the EPRL theory. The booster functions are responsible for the interference, which appear in the EPRL vertex amplitude and are not present in the topological one. They encode the imposition of the simplicity constraints in the theory. This interpretation agrees with the results in \cite{Dona:2018pxq}. Neglecting this interference results in a divergent upper bound estimate identical to the topological model.

Consider the Ponzano Regge model, a simpler spin foam theory that describes euclidean quantum gravity in three dimensions. The vertex renormalization amplitude in that theory (the 1-4 Pachner move) is cubically divergent in the cutoff $K^3$. The divergence is related to a residual gauge invariance in the path integral that is not entirely fixed \cite{Freidel:2002dw}. Geometrically it can be interpreted in the $\infty^3$ ways we can divide a tetrahedron in four with an extra point. The convergence of the EPRL vertex renormalization amplitude could signal that a similar symmetry is not present in the theory. The restriction to Lorentzian geometries with space-like boundaries of EPRL vertex in the large spins regime breaks the BF action's shift symmetry. Whether or not it indicates that simplicity constraints are imposed correctly in the EPRL theory remains an open question. We hope that a detailed analytical study and the contribution of other upcoming numerical techniques tailored to the study of the large spin limit of the theory \cite{Han:2021kll, Han:2023cen} could help solve this mystery. 

%%%%%%%%%%%%%%%%%%%%%%%%%%%%%%%%%%%%%%%%%%%%%%%%%%%%%%%%%%%%%%%%%%%%%%%%%%%%%%%%%%%%%%%%%%

\section{Acknowledgments}
P.D.~is supported by the ID\# 62312 grant from the John Templeton Foundation, as part of the ``The Quantum Information Structure of Spacetime (QISS)'' Project (\href{qiss.fr}{qiss.fr}).

\noindent P.F.~is supported by the Natural Science and Engineering Council of Canada (NSERC) through the Discovery Grant ``Loop Quantum Gravity: from Computation to Phenomenology''.

\noindent We also acknowledge the Shared Hierarchical Academic Research Computing Network (SHARCNET) and the Digital Research Alliance of Canada (\href{https://alliancecan.ca/en}{www.computecanada.ca}) for granting access to their high-performance computing resources. 

\noindent P.D. thanks Laurent Freidel for an insightful discussion on the consequence of the convergence of the EPRL vertex amplitude. P.F. thanks Francesco Gozzini for the useful discussions about this project.

\noindent We acknowledge the Anishinaabek, Haudenosaunee, L\=unaap\'eewak and Attawandaron peoples, on whose traditional lands Western University is located.

%%%%%%%%%%%%%%%%%%%%%%%%%%%%%%%%%%%%%%%%%%%%%%%%%%%%%%%%%%%%%%%%%%%%%%%%%%%%%%%%%%%%%%%%%%

\begin{appendices}

%%%%%%%%%%%%%%%%%%%%%%%%%%%%%%%%%%%%%%%%%%%%%%%%%%%%%%%%%%%%%%%%%%%%%%%%%%%%%%%%%%%%%%%%%%

\section{Vertex Amplitudes}
\label{app:vertexdetails}
In this Appendix, we report the definition of the Topological BF SU(2) spin foam vertex amplitude. 
\begin{equation}
    \label{eq:vertexBF}
    A_{v}^{BF}\left(  j_1, j_2, j_3, j_4 , j_5, j_6, j_7,  j_8, j_9,  j_{10} ; i_1 , i_2, i_3, i_4, i_5 \right) =
            \left \{ \begin{array}{ccccc}     
            i_{1} & j_{3} & i_{4} & j_{6} & i_{2} \\
            j_{4} & j_{10} & j_{8} & j_{5} & j_{1} \\ 
            j_{7} & i_{5} & j_{9} & i_{3} & j_{2}
        \end{array}\right \} \ .
\end{equation}  
The SU(2) invariant in \eqref{eq:vertexBF} is a $\{15j\}$ symbol of the first kind. According to the conventions of \cite{Yutsis:1962}, we write it in terms of Wigner's $\{6j\}$ symbols.
\begin{equation}
    \label{eq:15jsymbol}
    \begin{split}
        \left \{ \begin{array}{ccccc} 
        j_1 & j_2 & j_3 & j_4 & j_5  \\  
        l_1 & l_2 & l_3 & l_4 & l_5  \\ 
        k_1 & k_2 & k_3 & k_4 & k_5 
        \end{array}\right \} = 
        (-1)^{\sum_{i=1}^5 j_i + l_i +k_i} \sum_x& (2 x +1) 
        \Wsix{j_1}{k_1}{x}{k_2}{j_2}{l_1} \Wsix{j_2}{k_2}{x}{k_3}{j_3}{l_2}\\ 
  & \times \Wsix{j_3}{k_3}{x}{k_4}{j_4}{l_3} \Wsix{j_4}{k_4}{x}{k_5}{j_5}{l_4} \Wsix{j_5}{k_5}{x}{j_1}{k_1}{l_5}  \ .
    \end{split}
\end{equation}      
In this work we also use the in-line notation for 
\begin{equation}
\label{eq:6jsymbol}
\{ 6j \} (j_1,j_2,j_3,j_4,j_5,j_6) = \Wsix{j_1}{j_2}{j_3}{j_4}{j_5}{j_6} \ .
\end{equation}
In the booster function decomposition, the Lorentzian EPRL vertex amplitude is defined as:
\begin{equation}
\label{eq:EPRL_vertex_amplitude}
    \begin{split}
    &A_{v}\left(  j_1, j_2, j_3, j_4 , j_5, j_6, j_7,  j_8, j_9,  j_{10} ; i_1 , i_2, i_3, i_4, i_5 \right) = \\
    &\sum_{l_f=j_f}^{\infty}\sum_{k_e} \left\lbrace \begin{array}{ccccc} 
    i_{1} & j_{3} & k_{4} & l_{6} & k_{2} \\
    j_{4} & l_{10} & l_{8} & l_{5} & j_{1} \\ 
    l_{7} & k_{5} & l_{9} & k_{3} & j_{2} 
    \end{array}\right\rbrace (2k_2 + 1) (2k_3 + 1) (2k_4 + 1) (2k_5 + 1) \\
  &\hspace{1.5cm}B_4^\gamma\left(j_{5} , j_{6}, j_{7}, j_{1}, l_{5}, l_{6}, l_{7}, j_{1}; i_{2},k_{2}\right) 
  B_4^\gamma\left(j_{8} , j_{9}, j_{2}, j_{5}, l_{8}, l_{9}, j_{2}, l_{5}; i_{3},k_{3}\right) \\
  &\hspace{1.5cm}B_4^\gamma\left(j_{10} , j_{3}, j_{6}, j_{8}, l_{10}, j_{3}, l_{6}, l_{8}; i_{4},k_{4}\right) 
  B_4^\gamma\left(j_{4} , j_{7}, j_{9}, j_{10}, j_{4}, l_{7}, l_{9}, l_{10}; i_{5},k_{5}\right) \ .
\end{split}
\end{equation}
The booster functions are one-dimensional integrals over the rapidity parameter $r$ of the reduced matrix elements, in the $\gamma$-simple unitary representation of $SL(2,\mathrm{C})$.
\begin{equation}
\begin{split}
  \label{eq:boosterdef}
  B_4^\gamma&\left( j_1,j_2,j_3,j_4, l_1,l_2,l_3,l_4 ; i,k\right) 
  =\\
  &\sum_{ p_f } 
  \left(\begin{array}{cccc} l_1 & l_2 & l_3 & l_4 \\ p_1 & p_2 & p_3 & p_4 \end{array}\right)^{(k)}
  \left(\int_0^\infty \dd r \frac{1}{4\pi}\sinh^2r \, \bigotimes_{f=1}^4 d^{\gamma j_f,j_f}_{l_f j_f p_f}(r) \right)
  \left(\begin{array}{cccc} j_1 & j_2 & j_3 & j_4 \\ p_1 & p_2 & p_3 & p_4 \end{array}\right)^{(i)} 
  \ .
\end{split}
\end{equation}
The expression for $d_{jlm}^{\gamma j , j}(r)$ has been written in \cite{Ruhl:1970fk, Speziale:2016axj}:
\begin{equation}\label{eq:dSL2C}
\begin{split}
    d^{(\gamma j,j)}_{jlp}(r) =&  
    (-1)^{\frac{j-l}{2}} \frac{\Gamma\left( j + i \gamma j +1\right)}{\left|\Gamma\left(  j + i \gamma j +1\right)\right|} \frac{\Gamma\left( l - i \gamma j +1\right)}{\left|\Gamma\left(  l - i \gamma j +1\right)\right|} \frac{\sqrt{2j+1}\sqrt{2l+1}}{(j+l+1)!}  \\
    & \times \left[(2j)!(l+j)!(l-j)!\frac{(l+p)!(l-p)!}{(j+p)!(j-p)!}\right]^{1/2} e^{-(j-i\gamma j +p+1)r} \\
    & \times \sum_{s} \frac{(-1)^{s} \, e^{- 2 s r} }{s!(l-j-s)!} \, {}_2F_1[l+1-i\gamma j,j+p+1+s,j+l+2,1-e^{-2r}] \ .
\end{split}
\end{equation}
where ${}_{2}F_{1}$ is the Gauss hypergeometric function. In \eqref{eq:boosterdef} we used a short notation for the $(4jm)$ Wigner symbols. They are the contraction of two Wigner $(3jm)$ symbols (the unique intertwiner of three $SU(2)$ representation) labeled by the virtual spin $k$:
\begin{equation}
\label{eq:4jm_symbol}
\Wfour{j_1}{ j_2}{ j_3} {j_4}{p_1}{p_2}{p_3}{p_4}{k}
 = \sum_{p_k} (-1)^{k-p_k} \Wthree{ j_1}{ j_2}{ k} {p_1}{p_2}{p_k} \Wthree{k}{ j_3}{ j_4} {-p_k}{p_3}{p_4} \ .
\end{equation}
The large quantum number limit of the booster functions \eqref{eq:boosterdef} has been studied in \cite{Dona:2020xzv}. They possess an appealing geometrical interpretation in terms of boosted tetrahedra.

%%%%%%%%%%%%%%%%%%%%%%%%%%%%%%%%%%%%%%%%%%%%%%%%%%%%%%%%%%%%%%%%%%%%%%%%%%%%%%%%%%%%%%%%%%

\section{Monte Carlo summation}
\label{app:MCSums}
Suppose we want to compute the sum 
\begin{equation}
\label{eq:template_sum}
    S = \sum_{j\in \mathcal{J}} a_j \ ,
\end{equation}
where $\mathcal{J}$ is a finite subset of $(\mathbb{N}/2)^F$ and $j$ is a multi-index. So far, we are just muddling the waters with a complicated notation to give a nod to the spin foam application. If $V_\mathcal{J} = |\mathcal{J}|$ is the cardinally of the set, we can always map it in the interval of natural numbers between $1$ and $V_\mathcal{J}$. The sum \eqref{eq:template_sum} is a fancy way to represent a sum over an integer index from $1$ to $V_\mathcal{J}$.

If $V_\mathcal{J}$ is very large, the numerical computation of $S$ can result in a highly resource-hungry task. Therefore, we want to use (discrete) Monte Carlo techniques to approximate its value. 

We define a uniform probability density function over the set $\mathcal{J}$ as $R_\mathcal{J}$. We assume that $R_\mathcal{J}$ is normalized to $1$. The probability associated to every element of $\mathcal{J}$ with $R_\mathcal{J}$ is $ 1/ V_\mathcal{J}$. The hypothesis that the set $\mathcal{J}$ is finite ensures that $R_\mathcal{J}$ exists.

The fundamental step towards the implementation of Monte Carlo is to interpret the sum \eqref{eq:template_sum} as the expectation value of the terms of the sum $a_j$ using $R_{\mathcal{J}}$
\begin{equation}
\label{eq:expect_value}
S = V_\mathcal{J} \sum_{j \in \mathcal{J}} \frac{a_j}{V_\mathcal{J}} =V_\mathcal{J} \, E\left[ a (R_\mathcal{J}) \right] \ .
\end{equation}
We use a discrete uniform probability distribution because we assume we do not know in advance which term of the sum $a_j$ is contributing the most. This information would allow sampling from a more efficient probability distribution using importance sampling Monte Carlo. Alternatively, it would be possible to use Markov Chain Monte Carlo methods, such as the Metropolis-Hastings algorithm. This approach has been applied recently in \cite{Frisoni:2022urv} to compute observables in spin foams with many boundary degrees of freedom.

\medskip

We approximate the expectation value in \eqref{eq:expect_value} using a sample of the set $\mathcal{J}$. We use the probability distribution $R_\mathcal{J}$ and randomly extract $N_{mc}$ elements from the set $\mathcal{J}$. We denote this set as $\mathcal{J}^{mc}$. Strictly speaking, $\mathcal{J}_{mc}$ is not a subset of $\mathcal{J}$ since it can contain elements more than once. The Monte Carlo estimate of $S$ is given by
\begin{equation}
\label{eq:MC_estimate_def}
    S^{mc} = \frac{V_\mathcal{J}}{N_{mc}} \sum_{j \in \mathcal{J}^{mc}} a_j \ .
\end{equation}
The law of large numbers ensures that the average of a large number of samples becomes closer and closer to the expected value as more samples are performed. Since $S^{mc}$ is a sample of the sum $S$ we have that 
\begin{equation}
\label{eq:law_large_numbers}
S^{mc} \xrightarrow{N_{mc}\to \infty} S \ .
\end{equation}
The sum $S^{mc}$ \eqref{eq:MC_estimate_def} is the Monte Carlo estimate of $S$. We consider a ``large" number of samples $N_{mc}$ as we are interested in a numerical approximation of \eqref{eq:expect_value}.

The amount of computational resources necessary to compute $S^{mc}$ scales with the number of samples $N_{mc}$ and not the size of the original set $V_\mathcal{J}$. The number of samples $N_{mc}$ is a parameter of the calculation that we can tune. To have a good approximation and save resources, we must find a balanced value for $N_{mc}$. 

%%%%%%%%%%%%%%%%%%%%%%%%%%%%%%%%%%%%%%%%%%%%%%%%%%%%%%%%%%%%%%%%%%%%%%%%%%%%%%%%%%%%%%%%%%

\section{Aitken extrapolation}
\label{app:Aitken}
The Aitken's delta-squared process or Aitken extrapolation \cite{Aitken:1927} is a numerical recipe used to accelerate the rate of convergence of a sequence. Suppose you have a convergent sequence $S_n$ with $S = \lim_{n\to \infty} S_n$. If $S_n$ converges linearly, namely 
\begin{equation}
    \lim_{n\to \infty}\label{eq:linearconvergence}
    \frac{|S_n - S|}{|S_{n-1}-S|} = \lambda \ ,
\end{equation}
with $0<\lambda<1$. Linear convergence means the sequence is closer to its limit by almost the same amount with every step. In this case, we can approximate $S$ starting from the approximate relation 
\begin{equation}
\frac{S_n - S}{S_{n-1}-S} \approx \frac{S_{n-1} - S}{S_{n-2}-S} \ ,
\end{equation}
and solve for $S$ to find
\begin{equation}
\label{eq:Aitken}
 S \approx A[S_n]\equiv \frac{S_{n}S_{n-2} - S_{n-1}^2}{S_n - 2 S_{n-1} + S_{n-2}} \ .
\end{equation}
The sequence $A[S_n]$ is the Aitken extrapolation of $S_n$ and converge to $S$ faster than linear (see the original paper \cite{Aitken:1927} or the book \cite{Sidi2003} for a proof) meaning that
\begin{equation}
    \lim_{n\to \infty}\label{eq:faster}
    \frac{|A[S_n] - S|}{|S_{n}-S|} = 0 \ .
\end{equation}
The requirement of linear convergence of $S_n$ is equivalent to asking that for $n$ large enough
\begin{equation}
S_n \approx S + C \lambda^n \ ,
\end{equation}
for some constant $C$ and for $|\lambda|<1$. This allows us to estimate $\lambda$ without knowing the limit $S$, as is often the case. In practice, if we knew the value of the limit $S$, we would not need to extrapolate $S_n$. The limit of differences
\begin{equation}
\label{eq:finitedifference}
\lim_{n\to\infty}\frac{S_n - S_{n-1}}{S_{n-1}-S_{n-2}} = \lambda \ .
\end{equation}
If the rate convergence of $S_n$ is of higher order (quadratic or more), we do not need to extrapolate as the convergence is already very fast. However, if the convergence is sub-linear, we should look for a different extrapolation technique. We refer to the appendix of \cite{Dona:2022dxs} for an explicit and simple example. 

%%%%%%%%%%%%%%%%%%%%%%%%%%%%%%%%%%%%%%%%%%%%%%%%%%%%%%%%%%%%%%%%%%%%%%%%%%%%%%%%%%%%%%%%%%

\section{Diagrams of the melonic self-energy and vertex renormalization spin foam amplitudes}
\label{app:details_diagrams}
\begin{figure}[H]
    \centering
    \includegraphics[scale=0.55]{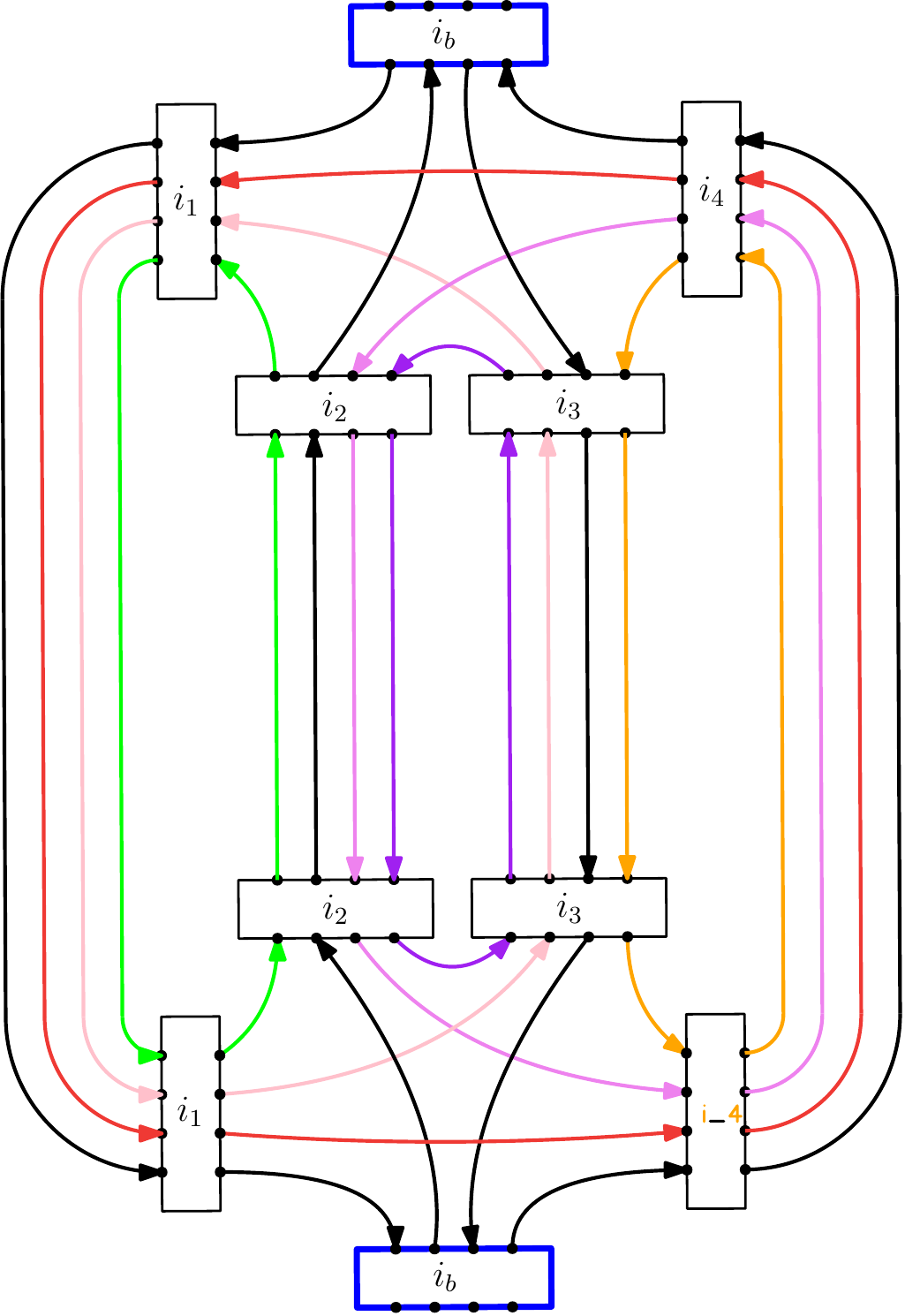}
    \caption{Wiring of the 2-complex of the self-energy spin foam diagram. The internal faces are highlighted with different colors. Boundary intertwiners have a blue box and also correspond to the integrals removed to regularize the amplitude.}
    \label{fig:2-complexes_wires}
\end{figure}

\begin{figure}[H]
    \centering
    \includegraphics[scale=0.55]{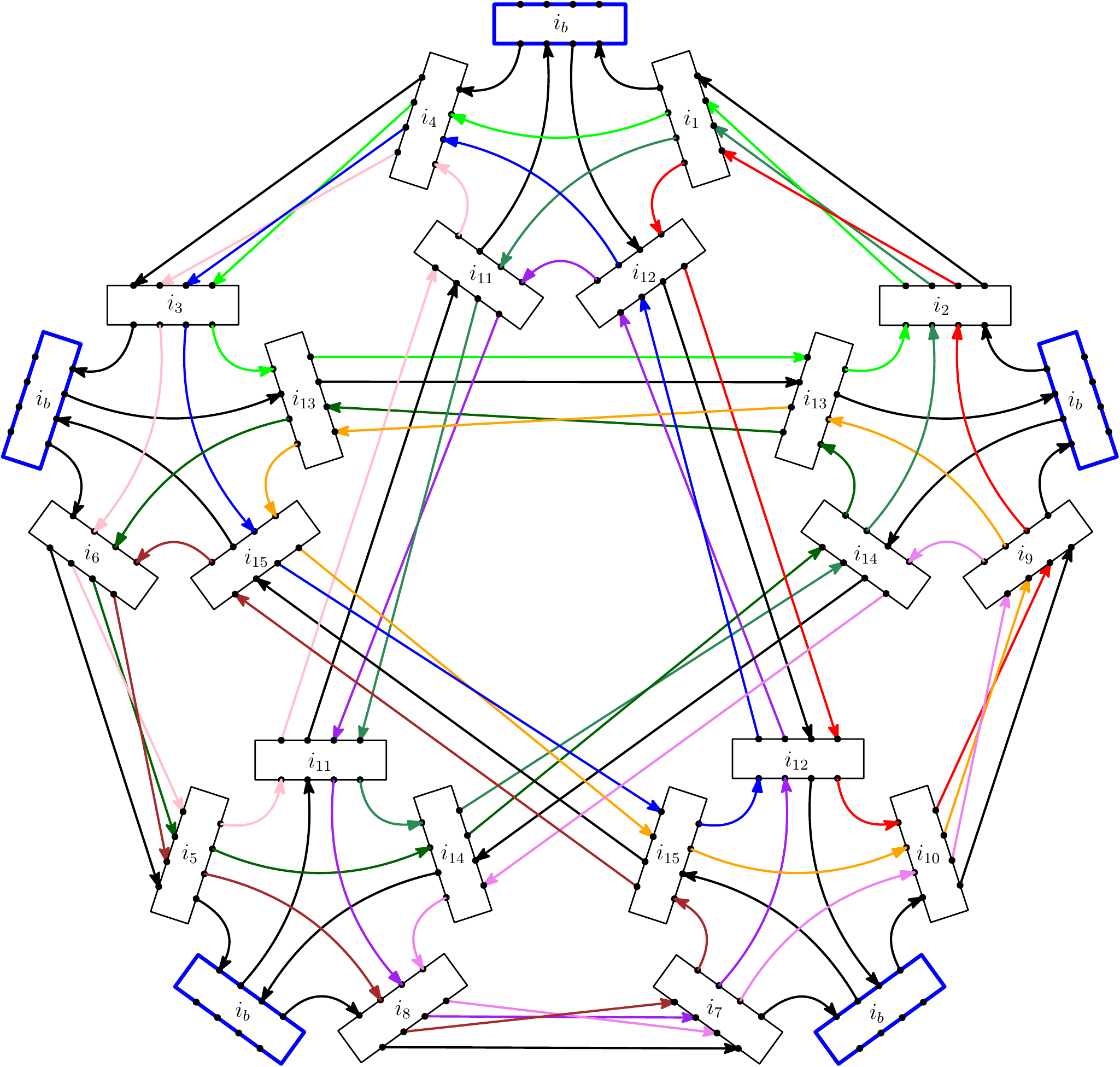}
    \caption{Wiring of the 2-complex of the vertex renormalization spin foam diagram. The internal faces are highlighted with different colors. Boundary intertwiners have a blue box and also correspond to the integrals removed to regularize the amplitude.}
    \label{fig:2-complexes_wires2}
\end{figure}

In this Appendix, we report the wiring diagrams of the self-energy and vertex renormalization spin foam amplitudes, highlighting the combinatorics of the internal faces. These are shown in Figure~\ref{fig:2-complexes_wires} and Figure~\ref{fig:2-complexes_wires2}. The notation for each intertwiner is used in equations \eqref{eq:melon} and \eqref{eq:vertex}. In order not to clutter the picture, we don't explicitly label the spins.

\end{appendices}

%%%%%%%%%%%%%%%%%%%%%%%%%%%%%%%%%%%%%%%%%%%%%%%%%%%%%%%%%%%%%%%%%%%%%%%%%%%%%%%%%%%%%%%%%%

\end{document}